\documentclass[manuscript,screen]{acmart}

\usepackage{booktabs} 
\usepackage{siunitx}
\usepackage{gensymb} 

\usepackage[ruled]{algorithm2e} 
\usepackage{url}
\usepackage{amsmath}
\usepackage{algorithmic}
\usepackage{amssymb}
\usepackage{todonotes}
\usepackage{soul}
\AtBeginDocument{%
  \providecommand\BibTeX{{%
    \normalfont B\kern-0.5em{\scshape i\kern-0.25em b}\kern-0.8em\TeX}}}

\newcommand\FF[1]{{#1}}
\newcommand\BB[1]{\textcolor{blue}{Bharath: #1}}

\begin{document}

\setcopyright{acmcopyright}
\acmJournal{TOSN}
\acmYear{2019} \acmVolume{1} \acmNumber{1} \acmArticle{1} \acmMonth{1} \acmPrice{15.00}
\title{ACES - Automatic Configuration of Energy Harvesting Sensors with Reinforcement Learning}

\author{Francesco Fraternali}
\email{frfrater@eng.ucsd.edu}
\affiliation{%
  \institution{University of California, San Diego}
  \city{San Diego}
  \country{USA}
}

\author{Bharathan Balaji}
\affiliation{%
  \institution{Amazon}
  \city{Seattle}
  \country{USA}}
\email{bhabalaj@amazon.com}

\author{Yuvraj Agarwal}
\affiliation{%
  \institution{Carnegie Mellon University}
  \city{Pittsburgh}
  \country{USA}
}
\email{yuvraj@cs.cmu.edu}

\author{Rajesh K. Gupta}
\affiliation{%
 \institution{University of California, San Diego}
 \city{San Diego}
 \country{USA}}
 \email{gupta@eng.ucsd.edu}

\renewcommand{\shortauthors}{Fraternali, et al.}

\begin{abstract}
Many modern smart building applications are supported by 
wireless sensors to sense physical parameters, given the flexibility they offer and the reduced cost of deployment. However, most wireless sensors are powered by batteries today and large deployments are inhibited by the requirement of periodic battery replacement. Energy harvesting sensors provide an attractive alternative, but they need to provide adequate quality of service to applications given the uncertainty of energy availability. We propose ACES, that 
uses reinforcement learning to maximize sensing quality of energy harvesting sensors for periodic and event-driven indoor sensing with available energy. Our custom-built sensor platform uses a supercapacitor to store energy 
and Bluetooth Low Energy to relay sensors data.
Using simulations and real deployments, 
we use the data collected to continually adapt the sensing of each node to changing environmental patterns and transfer learning to reduce the training time in real deployments. 
In our 60 node deployment lasting two weeks, nodes stop operations for only 0.1\% of the time, 
and collection of data is comparable with current battery-powered nodes. 
\FF{We show that ACES reduces the node duty-cycle period by an average of 33\% compared to three prior reinforcement learning techniques, while continuously learning environmental changes over-time.}
\end{abstract}



\maketitle

\section{Introduction}


Buildings are an essential part of modern society and benefit immensely from the Internet of Things (IoT) technologies~\cite{paper:IOT_directions}. Networked sensors are the bedrock of modern building services such as security, fire safety, energy, and lighting. Pervasive sensors deployed across a modern building can sense temperature, light, smoke, occupancy, energy use; it is possible to have hundreds to thousands of sensors in a modern office building \cite{khan2011big}. While traditional buildings use wired sensors, wireless technology is being increasingly adopted due to lower deployment cost and flexibility of placement~\cite{link:retrofitting-cost}. Much of the wireless sensors in the market are battery powered \cite{link:microDAQ-battery} and manual battery replacement is a key limitation that inhibits large scale deployments \cite{link:iot}. Energy harvesting sensors provide an attractive alternative, but their design needs to ensure adequate Quality of Service (QoS) given the limited and often uncertain energy availability. Many innovative battery-free solutions have therefore been proposed in prior works \cite{paper:campbell_cinamin,paper:campbell_energy_architecture_building, paper:ubicomp-2, Battery-free-cellphone, Pible}.  

Energy harvesting systems have to make a careful trade-off across sensing, communication, and computation requirements to maximize application utility with available energy~\cite{hsu2006adaptive, paper:RL-adapting}. These design tradeoffs change depending on hardware, application requirements and energy availability in the environment. Prior works either perform manual configuration or use heuristics to identify the operating points \cite{link:echelon, link:utc, link:daintree-PIR, link:enocean, Pible}. For example, EnOcean launched a commercial energy harvesting sensor in June 2019 that uses \textit{``a simple user interface consisting of one button and one LED allows for simple configuration without additional tools''}~\cite{link:enocean}. In our prior work, we designed the Pible platform\cite{Pible}, with the goal of harvesting enough energy to last an entire  day of operation. However, the key limitation of Pible is that its operational configuration was static and configured manually. It was, therefore, unable to adapt each sensor to the different environmental conditions that affect the amount of energy harvested. Manual configuration does not scale and heuristics do not generalize well to every context.



To overcome this limitation, we present our system \emph{ACES} that uses Reinforcement Learning (RL) ~\cite{sutton1998reinforcement} to automatically optimize the operation of sensors nodes to maximize QoS under uncertain energy availability.
In RL, an agent interacts with an environment and learns to make intelligent decisions with experience. A domain expert identifies the objective (i.e. reward function in RL terminology) and the inputs (i.e., state) that affect the decisions (i.e. actions) of the agent. The agent tries out different actions and learns from the feedback (rewards) received. RL algorithms are good at learning sequences of actions that maximize the long term expected cumulative rewards.
Several reasons make RL an appropriate solution for this problem \cite{paper:RL-adapting, paper:RL-ener-harvest-6797906, Scaling_my_ENSsys}: (i) \textit{automation}, by learning \textit{directly from experience}, RL can offer an alternative to heuristic-based approaches. An agent starts by knowing nothing about the environment and the task to be performed, but learns to make better decisions by interacting with the environment; (ii) \textit{optimization}, in \cite{mao2016resource} \FF{authors show} that RL is comparable or better than ad-hoc \FF{heuristics.} 
Given these attributes, RL is the perfect candidate to address the challenges presented earlier since we want a system that can detect environmental patterns (i.e. human presence, indoor lighting) whenever present, and automatically configure sensors over longer periods of time.


Hence, in case of energy harvesting, the long term objective ensures that it learns patterns in energy availability across days, nights and weekends. Energy availability is not perfectly predictable, especially in indoor conditions. The RL agent learns a strategy 
and is robust to noise (e.g. due to varying energy availability). 
However, learning an RL policy by directly trying out different actions in the real world can be time-consuming, convergence can take several weeks and exploratory actions can lead to poor QoS during training. ACES exploits a simulator that uses data from actual deployments to reduce convergence time and show that the policies learned by the simulator work demonstrably well once deployed in the real world at scale. Using an Intel Core-i7 CPU clocked at 3 GHz, our algorithm converges in <30 minutes of wall-clock time, 
making it feasible to use ACES in the real world for applications in which information dynamics vary hourly/daily. 

We implemented and evaluated ACES in a real deployment of energy harvesting sensors in our Computer Science Department building at UC San Diego. 
We built a Bluetooth Low Energy (BLE) based energy harvesting sensor platform, called Pible \cite{Pible}, that uses a solar panel to harvest energy and a supercapacitor to store energy. We use the node to sense light periodically and detect motion events with a PIR sensor. In smart buildings, having powered base station relays is typical \cite{paper:yuvraj_1}. The node sends the data to the closest base station, hence in this work, we only target one-hop sensor networks. 
We categorize 5 types of indoor lighting conditions and deploy a sensor node for each location type. We collect measurements such as light intensity and supercapacitor voltage level. Using these data traces, we develop a simulator that models the essential aspects of our sensor-node in different environments. We use the simulator to train the ACES RL agent using the Q-learning algorithm \cite{watkins1989learning}. 
We then perform multiple deployments for subsequent evaluations 
that traded off learning a one-time policy from historical data versus learning a new policy each day to better detect environmental diurnal patterns. 
Our real-world deployments show that ACES effectively learns from observing environmental changes and appropriately configures each sensor node to maximize sensing quality while avoiding energy storage depletion. 

To the best of our knowledge, ours is the \textit{first real-world deployment} that demonstrates an RL based sensing mechanism for energy harvesting sensors in indoor environments. 
We deploy 5 nodes, each exposed to different lighting conditions for 15 days, and use the light measurements collected to learn a policy in our simulator. We then deploy the learned policy in the same nodes 
and let the agent take action for 31 days. The nodes achieved a mean sampling period of 56 seconds, opportunistically collecting up to 1.7x more data than battery-powered sensors when energy is available and had 0\% dead time across nights and weekends. However, a one-time learning of a policy requires a data collection phase and can be susceptible to changes in the environment. Hence, we introduce a new training strategy that \textit{does not require any historical data}. We start with a default periodic policy on the first day and then train a new policy each day using data collected from all the previous days. 
This `day-by-day' training strategy learns a stable policy within a few days and achieves near-zero dead time. To alleviate poor performance in the initial training phase of a few days, we further reduce training time with transfer learning by training an initial policy using data collected by sensor nodes in 5 different lighting conditions.  
Finally, we generalize ACES formulation to event-driven sensing. We deployed 45 nodes with PIR event sensing and periodic light sensing for two weeks. ACES nodes could detect 86\% of the events on average compared to a baseline of battery-powered nodes which detected all events. We expanded our deployment to 15 more nodes with periodic light sensing to validate the findings of our initial 5 node deployment. 


We compare ACES with three RL Q-learning based methods proposed in prior works and a local-mote heuristic algorithm. In a week-long simulation experiment, we show that \FF{ACES reduces the node duty-cycle period by 33\% on average} w.r.t. prior RL techniques, while continuously learning environmental changes over-time.  
ACES is open source, and all the software is available online \footnote{https://github.com/francescofraternali/ACES.git}.

\section{Related Work}
\label{RelWork}
A multitude of innovative energy harvesting solutions has been proposed in the literature~\cite{sudevalayam2011energy, ulukus2015energy, paper:Monjolo, naderiparizi2015wispcam}. In several works,
 the sensors use up energy as it becomes available ~\cite{paper:campbell_energy_architecture_building, campbell2014energy, naderiparizi2015wispcam,paper:Monjolo}, e.g., harvesting AC power lines for energy metering~\cite{paper:Monjolo}, RFID based battery-free camera~\cite{naderiparizi2015wispcam}, thermoelectric harvesting based flow sensor~\cite{paper:DoubleDip}. Backscatter sensors are a special case that eliminate communication-based energy expenditure by using existing RF signals for both communication and harvesting~\cite{wifi-backscatter,ble-backscatter,Battery-free-cellphone}. However, sensing whenever energy is available is not ideal for all applications - sensors may miss important events because they do not store enough energy or send too much data when it is not needed~\cite{paper:iot-powering, paper:data-energy-tradeoff}. A solar panel powered temperature sensor should work on nights and weekends even when energy availability is low and a backscatter-based motion sensor should capture events when there are no RF transmissions. Hence, these sensors nodes need to be carefully configured to increase their quality of service and compete with existing battery-based solutions. 
 
 Other related work tackles the intermittent operation of battery-less energy harvesting systems themselves \cite{hester2017flicker, Hester:2017:TEI:3131672.3131673, Maeng:2017:AIE:3152284.3133920alpaca, lucia2017intermittent}. In contrast, our work is looking to solve an orthogonal problem, where we are optimizing our system to perform sensing, computation, and communication while managing limited resources. 

Reducing manual intervention for sensor configuration is an important task as underlined by many works~\cite{paper:IoT-reconfigurable,paper:direct-RL, paper:sens-average,paper:RL-adapting,paper:RL-energy-harvesting,paper:RL-ener-harvest-QoS-5284227,Pible}. 
Adaptive duty cycling of energy harvesting sensors has been used to achieve energy-neutral operations \cite{paper:Mani-power-management, hsu2006adaptive, paper:sens-average,vigorito2007adaptive}: nodes adjust their duty-cycle parameters based on the predicted energy availability to increase lifetime and application performance \cite{sudevalayam2011energy}. 
Moser et al.~\cite{paper:sens-average} adapt parameters of the application itself to maximize the long-term utility based on future energy availability prediction. Reinforcement learning (RL) also predicts energy availability, but unlike the above solutions, it also learns an optimal policy that maximizes the long-term reward. Hence, it makes better decisions compared to utilizing heuristics as we show in our evaluation.

Several works have the goal to maximize the overall Quality of Monitoring (QoM) in wireless rechargeable sensor networks (WRSNs) \cite{dai2013practical, Yau:2010:QMS:1824766.1824774, wei2018reinforcement}. Authors in \cite{dai2013practical, Yau:2010:QMS:1824766.1824774} investigate the sensor scheduling problem aiming to maximize QoM for events capture. They solve a much harder problem: detecting events with a collection of sensors. While their work provides theoretical guarantees, it either assumes events are completely random or have specific distributions. Results are based on simulations and \FF{authors did not consider noise variability in real-world data}. We focus on a pragmatic solution for a collection of independent sensing nodes for periodic sensing with harvesting. Our work is experimental, where we use real data and real sensors in a building. Our RL algorithm is agnostic to the event distribution as the agent learns the distribution of events based on past data. We have solved a number of issues in the deployment of RL in practice with day-by-day learning and transfer learning.

Zhenchun et al. \cite{wei2018reinforcement} use RL to autonomously plan and adjust the charging path of mobile chargers, to charge more efficiently in a sensor network. Our goal is orthogonal to this problem, as we only rely on solar-based charging and do not need mobile chargers. We show that even without using batteries we can achieve almost 0\% dead time, making it a much simpler and cheaper solution than \cite{wei2018reinforcement}.

RL has been identified by several prior efforts as a mechanism to configure wireless sensors and has shown promising results based on simulations~\cite{paper:RL-review, RLMan, paper:RL-adapting, paper:RL-ener-harvest-6797906}. 
Zhu et al.~\cite{paper:RL-ener-harvest-6797906} apply RL for sustaining perpetual operations and satisfying the throughput demand requirements for energy harvesting nodes (i.e. RLTDPM). However, their algorithm is built and tested in an outdoor environment, where sunlight patterns are consistent throughout the day (light is available from sunrise to sunset). We focus on indoor sensing where the daylight patterns are less well-defined and the light availability is affected by human occupancy.
With respect to our work, RLTDPM is a one-time learning, and even if it can adjust its sensing to application's requirements, it is not able to learn changes in the environment over-time.
Shaswot et al. \cite{adaptive-power-management-RL} use solar energy harvesting sensor nodes powered by a battery to teach an RL system to get energy-neutral operations. They use the SARSA algorithm~\cite{sutton1998reinforcement} to study the impact of weather, battery degradation and changes to the hardware. However, their work is also limited to simulations of outdoor environments.

Simulation results by Dias et al.~\cite{paper:RL-adapting} optimize energy efficiency based on data collected over five days from five sensor nodes using on-line reinforcement learning with Q-Learning (i.e. On-Line RL). However, their reward function does not depend on battery level or energy consumed, and thus, does not capture realistic conditions.
We compared ACES with On-Line RL, showing that our system can reduce nodes's duty-cycle period by up to 35\%. 
They also assume a fixed 12 hours period as the time needed by the Q-Learning algorithm to learn the action-value function for the rest of the 4.5 days experiment. However, a one-time training scheme is unable to capture all kinds of environmental changes. In our work, we propose a periodic training method that maximizes system performance even in the presence of environmental changes.

Aoudia et al. present RLMan \cite{RLMan} that uses the actor-critic algorithm with linear function approximation~\cite{grondman2012survey}. They use existing indoor and outdoor light measurements for simulations. While the authors claim their linear function approximations facilitate RL training in a wireless node, they do not deploy RL on a real node, nor do they report their memory and compute requirements. 
Our formulation performs RL training on a local server and eliminates the need for putting RL inside the node. We use the Q-learning algorithm, and our Q-table is only 25kB in size and demonstrate that it can be embedded in the sensor node as well. 
Our prior work~\cite{Scaling_my_ENSsys} used the Q-learning algorithm for adapting the sampling rate. The paper focused on methods to scale RL to large deployments and reduce training time. Using simulations, we showed that training a new policy periodically based on real light data improved the adaptability to changing conditions. We also showed that transfer reinforcement learning works well and it is possible to share the learned policies among sensor nodes that share similar lighting conditions. However, all of our results were based on simulations from 5 sensor nodes in multiple lighting conditions and were not deployed or tested in the real world. In this work, we show that these results transfer to the real-world with multiple large scale deployments. We have also extended our ACES RL formulation to event-driven sensors and evaluate their performance in real deployments. Besides, we analyze the effect of changing the state and action space and quantify the energy neutrality of the learned policies. 


To the best of our knowledge, our work is the first to exploit reinforcement learning for adaptive sampling in energy harvesting nodes in a \emph{real deployment}. The RL design needs to take into account changing environmental conditions, imprecise state estimates, and stochastic state transitions. We present a formulation of our problem and perform simulation-based modeling to capture realistic conditions that transfer to the real world. 
Furthermore, we have extended our problem formulation to accommodate \emph{event-driven sensors} such as PIR motion detectors. The sensor node needs to preserve enough energy to communicate events as they occur. We show that ACES successfully learns typical event firing patterns and curtails its sensing period to account for the additional energy use.
\begin{table}[ht]
\centering
  \caption{Features Comparison of ACES w.r.t. Prior Techniques}
  \label{tab:Related-work}
  \begin{tabular}{c c c c c}
    \toprule
     & Learn & Deployment & Algorithm Execution & Tested in\\
     & Over-Time & Time & in the Node &  Real-World\\
    \midrule
    Mote-Local \cite{Pible} & No & $\sim$ 0 & Yes & Yes\\
    RLMan \cite{RLMan} & No & Days & Yes & No\\
    On-Line RL \cite{paper:RL-adapting} & No & Hours & No & No\\
    RLTDPM \cite{paper:RL-ener-harvest-6797906} & No & Days & No & No\\
    WRSN \cite{dai2013practical} & No & $\sim$ 0 & No & No\\
    ACES & Yes & $\sim$ 0 with TL & Yes & Yes \\
  \bottomrule
  \multicolumn{5}{c}{Legend: TL = Transfer Learning}\\
  
\end{tabular}
\end{table}

Table \ref{tab:Related-work} reports the main differences between current state-of-the-art techniques including three RL methods and ACES. As we can see, ACES is the only work that is tested in the real-world while using RL. It overcomes typical real-world problems such as continuously adapting to new changes in the environment and reducing deployment-time to almost zero due to transfer learning. Furthermore, ACES can limit system failures due to network disconnections by executing the learned algorithm inside the node (Section \ref{sec:failures}).

\section{ACES Design and Implementation}
\label{Design}


\subsection{Problem Statement}
In buildings, sensors are used for applications such as environment control (e.g. air conditioning), safety, security or convenience. We broadly categorize the sensing as periodic (e.g light intensity sensor) or event-driven (e.g. motion sensor). For periodic sensors, the higher the frequency, the more responsive the control systems and better the QoS. \FF{For event-driven sensors, reducing the number of missed events translate to higher QoS.} However, the QoS is extremely poor if the sensors are not operational for hours at a time - control systems will lose their feedback loop, event-driven applications will be non-functional. Hence, we define our objective as to minimize the average duty cycle period while ensuring the energy harvesting nodes remain alive. \FF{As an example, a sensor with an average duty cycle period of 10 seconds w.r.t one with a 50 seconds duty cycle period, will send 5 times more data.} 


Nodes can be placed in different locations in a building. Each sensor node will be subject to different lighting patterns that are determined by human behavior (e.g. lights turned on and off) or by natural light whenever the node is close to a natural source of light (e.g. window). Light availability will vary from weekdays to weekends, from winter to summer and with changes in usage patterns, e.g. a conference room vs a lobby. 
The sensor node needs to adapt itself to these changing conditions to maximize the utility of its applications. A node placed near a window and subject to a source of natural sunlight can still be considered as an "indoor" case for two main factors: \FF{(i) indoor light can still impact the energy gathered}, (ii) blinds are common and reduce outside lighting.

\subsection{Reinforcement Learning and Q-Learning}
\label{RL}

In a typical RL problem \cite{sutton1998reinforcement}, an agent starts in a state $s$ and by choosing an action $a$, it receives a reward $r$ and moves to a new state $s^\prime$. This process is repeated until a final state is reached. RL agent's goal is to find the best sequence of actions that maximizes the long term reward. 
The way the agent chooses actions in each state is called its policy $\pi$.
\begin{equation}
s \overset{a}{\underset{\pi}\rightarrow} r, s^\prime
\end{equation}
For each given state $s$ and action $a$, we define a function $Q(s, a)$ that returns an estimate of a total discounted reward we would achieve by starting at state $s$, taking the action $a$ and then following a given policy $\pi$ till a final state is reached. 
\begin{equation}
\label{eq:recursive}
Q_\pi(s, a) = r_0 + \gamma r_1 + \gamma^2 r_2 + \gamma^3 r_3 + ... 
\end{equation}
where $\gamma \leqslant 1$ is called a discount factor and it determines how much the function $Q$ in state $s$ depends on the past actions (i.e. how long in the past does the agent see) since each member in the equation exponentially diminishes the further they are in the past. 
Equation (\ref{eq:recursive}) can be rewritten in a recursive form called the Bellman equation: 
\begin{equation}
\label{eq:gen_opt}
Q_\pi(s, a) = r_0 + \gamma (r_1 + \gamma r_2 + \gamma^2 r_3 + ...) = r_0 + \gamma max_a Q_\pi(s', a)
\end{equation}
 The Q-learning algorithm starts with a randomly initialized Q-value for each state-action pair and an initial state $s_0$. An episode is defined as a sequence of state transitions from the initial state to the terminal state. The algorithm follows a $\epsilon$-greedy policy, where for each state it picks the action that has the maximum Q-value with probability $(1-\epsilon)$ and a random action otherwise. The reward obtained by selecting the action is used to update the Q-value with a small learning rate. Under the conditions that each of the state-action pairs are visited infinitely often, the Q-learning algorithm is proven to converge to the optimal function $Q^*$~\cite{watkins1992q}. The optimal policy $\pi^*$ takes the action that maximizes $Q^*$ in each state. The $\epsilon$-greedy policy is used as an exploitation exploration trade-off~\cite{sutton1998reinforcement}, where the occasional random actions encourage the agent to explore the state-action space. It is typical to use a high value of $\epsilon$ at the start of training and gradually reduce it over time to increase exploitation.  
 
 Between different RL algorithms, we choose Q-learning because it 
 is easy to use when there is a set of discrete states and actions. Q-learning is an off-policy algorithm, which means we can learn a policy (and a Q function) with historical data when available. By exploiting the off-policy nature of Q-learning, we introduce two additional variants to the algorithm to reduce convergence time: day-by-day learning and transfer learning. In day-by-day learning, we learn a new Q function each day based on data collected in the past day. The Q function gradually improves each day and can accommodate changing environmental conditions. With transfer learning, we use a Q function learned for another sensor node as an initial policy for a new node. 
 
Q-learning is guaranteed to converge to an optimal solution under certain conditions: (1) we accurately know the dynamics of the environment; (2) we have enough computational resources to complete the computation of the solution; and (3) the Markov property. Hence, in our case, Q-learning is not guaranteed to converge to an optimal solution since as we show in Section \ref{sec:problem} there are a number of problems such as: 
(1) we discretize the state/actions and make assumptions on the energy consumption of the components; (2) we have limited computational resources and training time. Therefore, there could be some states that are not seen during training. Even with these approximations, we show that our system can find a solution that outperforms state-of-the-art results.

\subsection{RL Problem Formulation for Energy Harvesting Sensors}
\label{sec:problem}

\begin{figure}
  \centering
     \includegraphics[width=0.4\linewidth]
{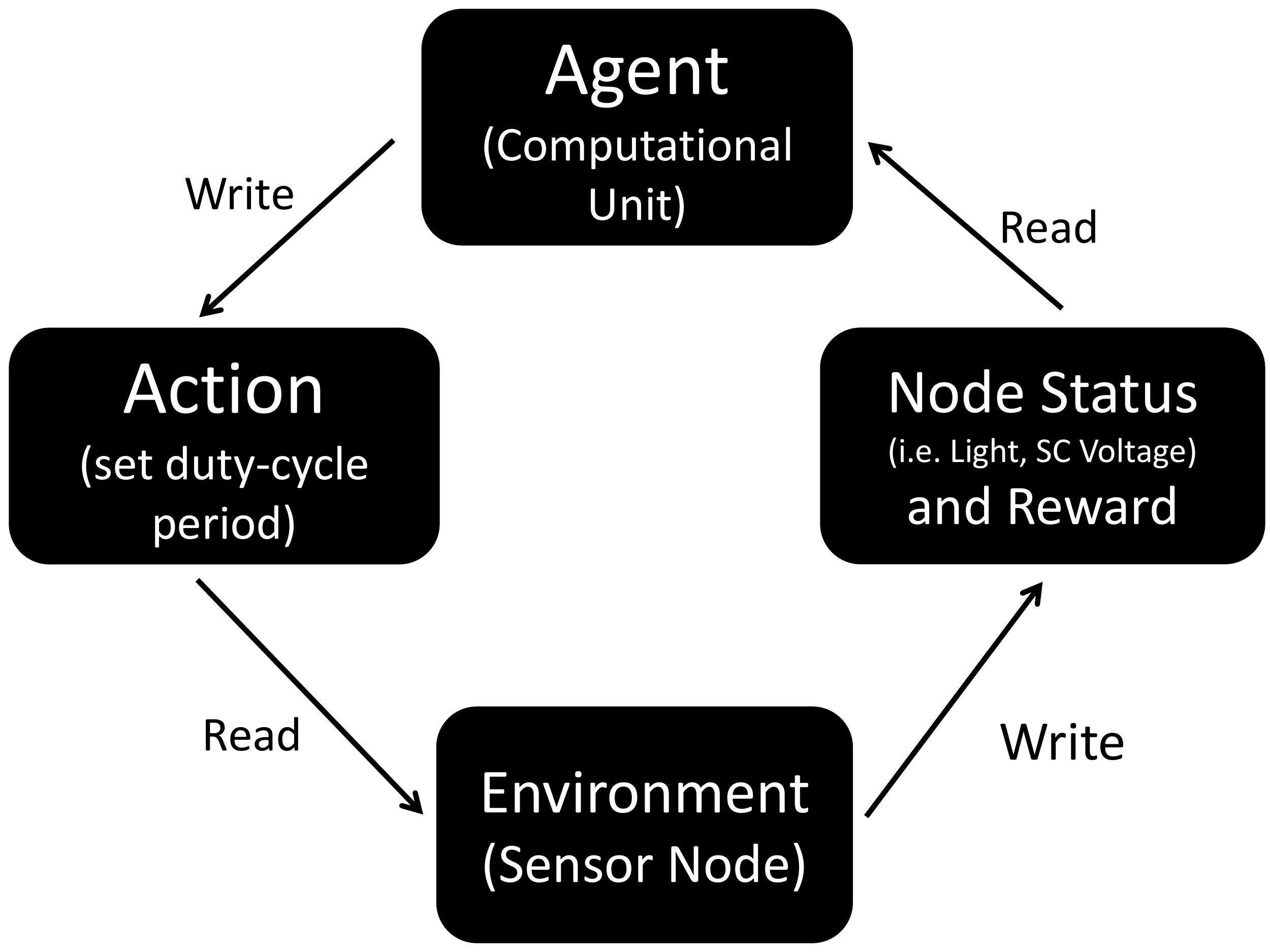}
    \caption{Reinforcement Learning Communication Process for Energy Harvesting Sensors: Block Diagram.} 
    \label{fig:Circle_Life}
\end{figure}

Figure \ref{fig:Circle_Life} shows an overview of our RL formulation for the configuration of energy harvesting sensor nodes. A sensor node collects energy from a solar panel and sends sensor data to a computational unit that runs the RL policy and determines the node sleep time (i.e. the time the node stays in sleep mode).
The computational unit can be the node, a base station, a local server or in the cloud as long as they have enough computational power to train RL policies. We run the training on the local server.


\textbf{Objective and Quality of Service (QoS):} the objective of our problem is twofold:
\begin{itemize}

\item[--] Maximize sensing (reduce duty-cycle period by reducing the node sleep time) so we capture periodic sensor measurements more frequently and minimize missed events for event-driven sensors. 

\item[--] Minimize dead time. If the sensor node dies, it enters a cold start phase and could not send data for hours.

\end{itemize}

Hence, we define the Quality of Service (QoS) as the duty cycle period averaged across a period of time (e.g. 7 days):
\begin{equation}
\label{objective}
\textrm{Duty-cycle-period = $\frac{1}{\tau} \sum_{t=0}^{\tau} $ T$_{sensing}$ + T$_{sleep}$}
\end{equation}

where, T$_{sensing}$ is a constant and includes the total time the node remains active (i.e. the time the node wakes-up, read sensor data, send it over BLE and goes back to sleep), T$_{sleep}$ is the node sleep time (i.e. MCU in sleep mode) and $\tau$ is the total number of duty cycles in a given period. For event-driven sensors (e.g. motion sensor), a sensor is left awake during T$_{sleep}$ until an interrupt is raised due to an event. The RL agent decides the T$_{sleep}$ period for both types of sensors. 
Another metric commonly used in literature is duty cycle ratio \cite{duty-cycle}. We can compute the duty cycle ratio using the proposed duty cycle period with the following equation:

\begin{equation}
\label{duty-cycle-ratio}
\textrm{Duty-cycle-ratio = $\frac{T_{sensing}}{\textrm{Duty-cycle-period}}$}
\end{equation}

\textbf{Agent:} The agent is the program in the computational unit that takes in measurements from the sensor, outputs the node sleep time (i.e. T$_{sleep}$) to use and updates the RL policy based on rewards.

\textbf{Environment:} 
It is everything outside the agent, which includes sensor nodes, the wireless channel, the lighting conditions and events that trigger the sensors. 


\begin{table}[h]
\centering
  \caption{Node Sleep Time based on Action Index. One of the objective of our system is to reduce the \textit{Duty-Cycle Period} of the system by reducing the Node Sleep Time. With \textit{Duty-Cycle Period} we indicate the time the node stays in sleep mode (i.e. T$_{sleep}$) plus the time it remains active (i.e. T$_{sensing}$) such as waking-up, reading the sensors and transmitting the data using a wireless protocol.}
  \label{tab:Perf-State}
  \begin{tabular}{cc}
    \toprule
    Action Index & Node Sleep Time [s]\\
    \midrule
    3 & 15\\
    2 & 60 (1 min)\\
    1 & 300 (5 min)\\
    0 & 900 (15 min)\\
  \bottomrule
  \vspace{-2em}
\end{tabular}
\end{table}

\textbf{State:} We use: \textit{(i)} light intensity, \textit{(ii)} energy storage level, \textit{(iii)} weekend/weekdays. We discretize light intensity and energy storage levels to 10 values each. 
The discretization helps reduce the state space, and hence, decreases the convergence time of the Q-learning algorithm.

\textbf{Action:} 
To change the duty cycle period, from \ref{objective}, we can only act on T$_{sleep}$ since T$_{sensing}$ is dependent on the sensing characteristics and the communication mechanism. 
Therefore, we discretize the node sleep time to meet typical commercial duty-cycle periods for IoT devices in buildings, since they are in the order of minutes but can range from tens of seconds to an hour \cite{paper:augment_audits-finnigan2017augmenting, link:sens-rate-1,link:sens-rate-2}. We report our node sleep time discretization in Table \ref{tab:Perf-State}. For periodic sensors, the action selects the node sleep time to use, e.g. action 2 corresponds to a sleep time of 1 minute. For event-driven sensors, we observe that once an event is triggered, a subsequent event within a short amount of time is inconsequential. Hence, we keep the node alive until an event occurs, after which it sends the event packet and sleeps for the period indicated by the action. \FF{Any event during sleep time is missed.}  

\textbf{State Transitions:} The agent observes state transitions and takes actions, i.e. sends commands to change the node sleep time, every 15 minutes. A small timestep for state transitions increases the communication overhead between the base station and sensor node and a large timestep misses the opportunity to tune the sensing rate in a fine-grained manner. We select a timestep of 15 mins as a design trade-off between these factors. We use 24 hours as our episode, starting when the node is turned on. We \FF{will} consider dynamic transitions based on node sleep time as future work.

\textbf{Reward Function:} The reward function is a trade-off between maximizing sensing that consumes energy while penalizing dead time. 
We assign rewards as follows:
\begin{itemize}
\item[--] Reward = Action index (i.e. 0, 1, 2 or 3)

\item[--] Reward = -300 if energy storage level reaches 0.

\end{itemize}
If the energy storage reaches 0, the system should receive a negative reward that dissolves all the benefit of using the maximum index action available (i.e. action 3) since it could cause a dead time lasting several hours.
Therefore, with the above reward function, the negative reward should be compensating for the maximum total reward in 24 hours. The following formula makes this calculation: 
\begin{equation}
\label{300formula}
Reward <= - (\textrm{number of state-transitions per day * max action}\  = 24 * 4 * 3 = 288)
\end{equation}

Using Formula (\ref{300formula}), we picked -300 as a negative reward. 
The general principle here is to identify the objective function of the problem and formulate it as a reward function. Any constraints, such as avoiding the depletion of energy in storage, can be expressed as a negative reward. While we use a heuristic (i.e. Formula \ref{300formula}) to identify the negative reward coefficient, one can also identify the coefficient automatically using Lagrangian relaxation~\cite{bohez2019value}. Q-learning requires the use of discrete states and actions, and the range of actions can be decided based on the desired application requirements. Fine-grained discretization can lead to a better policy, but increases convergence time. We study design trade-off between different discretizations in Section \ref{sec:discrete}. Our problem formulation can be easily adapted to different sensors, energy harvesters, and applications.

\label{sec:rl_circle}
\begin{algorithm}[ht]
  \caption{RL Algorithm for Energy Harvesting Sensors}
  \label{algoRL}
  \begin{algorithmic}[1]
  \STATE Initialize q$_{table}$ as an empty set
  \STATE Initialize control action a, state s$_{curr}$, s$_{next}$ \textit{time passed} = 0
  \STATE $\epsilon$ = $\epsilon_{min}$
  \STATE $s_{curr}$ $\leftarrow$ Sense Environment 
  \WHILE{\textit{time passed} $<$ \textit{episode\_duration}}
  
  \STATE \newenvironment{rcases}
  {\left\lbrace\begin{aligned}}
  {\end{aligned}\right.}

\begin{equation*}
\text{ a = }
\begin{rcases}
  \text{if (uniform(0,1) <= } \epsilon \text{)   take:   }  argmaxQ(s_{curr}, a')\\
  \text{otherwise:    }  \text{    take a random action}
\end{rcases}
\end{equation*}
\STATE wait for \textit{T} time units, \textit{time passed} += \textit{T}
\STATE $s_{next}$ $\leftarrow$ Sense Environment 

\STATE r = reward(s$_{curr}$, a, s$_{next}$)

\STATE $ q_{predict} = q_{table}$[$s_{curr}, a$]$ $
\STATE $ q_{target} = r + \gamma * max(q_{table}$[$s_{next},$ \FF{$\forall$ a}])
\STATE $ q_{table}[s_{curr}, a] += \alpha * (q_{target} - q_{predict}) $

\STATE $ \epsilon_{min} = \epsilon_{min} + \Delta $

\STATE $ \epsilon = min(\epsilon_{max}, \epsilon_{min})$
\STATE $ s_{curr} \leftarrow s_{next} $
  \ENDWHILE
  \end{algorithmic}
  The specific values of each variable used are given in Table \ref{Design}.
\end{algorithm}


\subsection{RL Algorithm}
Algorithm \ref{algoRL} details the use of Q-learning in ACES. After initialization (line 1-3), the agent receives the node state - light, energy storage voltage and current action index (line 4). The algorithm starts the episode and selects an action following the $\epsilon$-greedy policy (line 6). In the beginning, $\epsilon$ is initialized to $\epsilon_{min}$, the algorithm selects a lot of random actions in the first phase of the learning since uniform() produces a random value uniformly distributed between 0 and 1. At each time step (15 mins), $\epsilon$ is increased by $\Delta$ (line 13-14), causing the action policy to select more exploitative actions over time.

The sensor node updates its node sleep time and sends its state again after \textit{T} time units (line 7-8). We calculate the rewards with the current state, action and next state (line 9). 
We assume our system has reached convergence when the mean of all the values in the Q-Table does not change their value by >5\%. 

We progressively improved our policy training strategy to reduce time to convergence.

\textbf{One-Time Learning:} We train a policy in a simulator based on sensor measurements collected for 15 days. This is a typical setting used in prior works~\cite{RLMan, paper:RL-adapting}.

\textbf{Day-by-Day Learning:} We train a policy every day based on the sensor measurements collected in the past 24 hours. Each day's training starts with the policy learned in the previous day. The policy converges within a few days to achieve energy-neutral operations. The daily training also adjusts the policy to changing environment conditions (non-stationary environments). This borrows from Batch RL methods in literature~\cite{kalyanakrishnan2007batch}.

\textbf{Transfer Learning:} Instead of learning a policy from scratch, the initial policy is borrowed from another node's converged policy. We achieve 0\% dead time from the first day of deployment with transfer learning. We continue to use the day-by-day learning procedure for iterative improvement of the policy.

\begin{figure*}[ht]
  \centering
     \includegraphics[width=\linewidth]{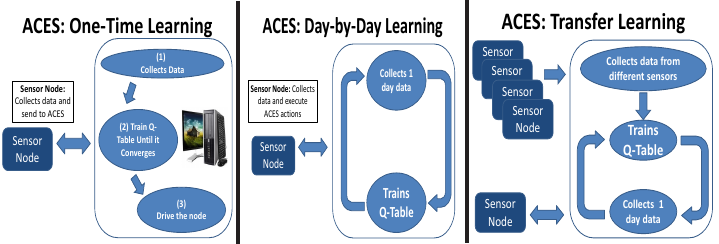}
    \caption{A comparative overview of the one-time, day-by-day and transfer learning methods used in ACES. Day-by-day learning assures continual learning to changing conditions and removes a separate data collection period. Transfer learning minimizes dead time during initial phase of the deployment.
    }
    \label{fig:Learn_algo}
\end{figure*}

Figure \ref{fig:Learn_algo} shows an overview of one-time, day-by-day and transfer learning methods. Day-by-day learning assures continual learning to changing conditions and removes a separate data collection period. Transfer learning \FF{reduces} dead time during initial phase of the deployment.

Table \ref{tab:HyperP-Simu} reports the hyper-parameters used for the Q-learning algorithm for our simulations.

\begin{table}[ht]
\centering
  \caption{Q-Learning Hyper-parameters used for simulations}
  \label{tab:HyperP-Simu}
  \begin{tabular}{c c}
    \toprule
    Hyper-Parameter & Value \\
    \midrule
    Reward-decay ($\gamma$) & 0.99 \\
    Epsilon max ($\epsilon_{max}$) & 1 \\
    Epsilon min ($\epsilon_{min}$) & 0.1\\
    Epsilon increment ($\Delta$) & 0.0004\\
    Learning rate ($\alpha$) & 0.1\\
    Episode Duration & 24 hours\\
    Wait Time \textit{T} & 15 mins\\
    Training phase & 15 days\\
  \bottomrule
\end{tabular}
\end{table}

\subsection{Hardware and Communication Process}

\subsubsection{Sensor Node:} 
\begin{figure}[ht]
  \centering
     \includegraphics[width=0.8\linewidth]{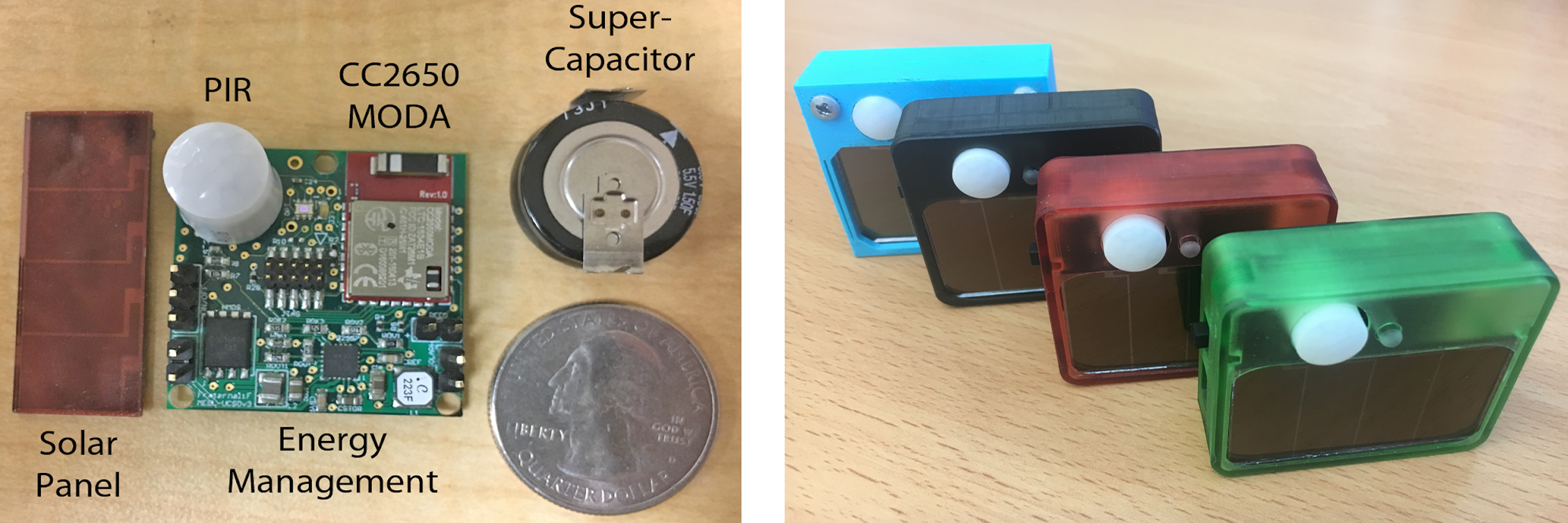}
    \caption{General Energy Harvesting Sensor-Node. 
    }
    \label{fig:Pible}
\end{figure}

We developed a general-purpose energy harvesting battery-less sensor (Figure~\ref{fig:Pible})~\cite{Pible, fraternali2018pible}. We use a solar panel (AM-1454 \cite{link:sanyo}) as our energy harvester and store energy in a super-capacitor \cite{link:supercap55}. Once the voltage reaches a usable level, an energy management board powers the microcontroller (MCU) \FF{to start its operations.} We use a 1F super-capacitor with a 5.5V nominal voltage. 
The MCU uses a high resistance voltage divider (10MOhm) to minimize leakage current. We use BLE for communication and the node has light and PIR sensors. The sensor-node achieves up to a week of lifetime without light when it sends one sensor measurement every 10 minutes.

\subsubsection{Wireless Sensor Network Architecture:}
\label{sec:WSN}
\begin{figure}[ht]
\centering
     \includegraphics[width=0.7\linewidth]{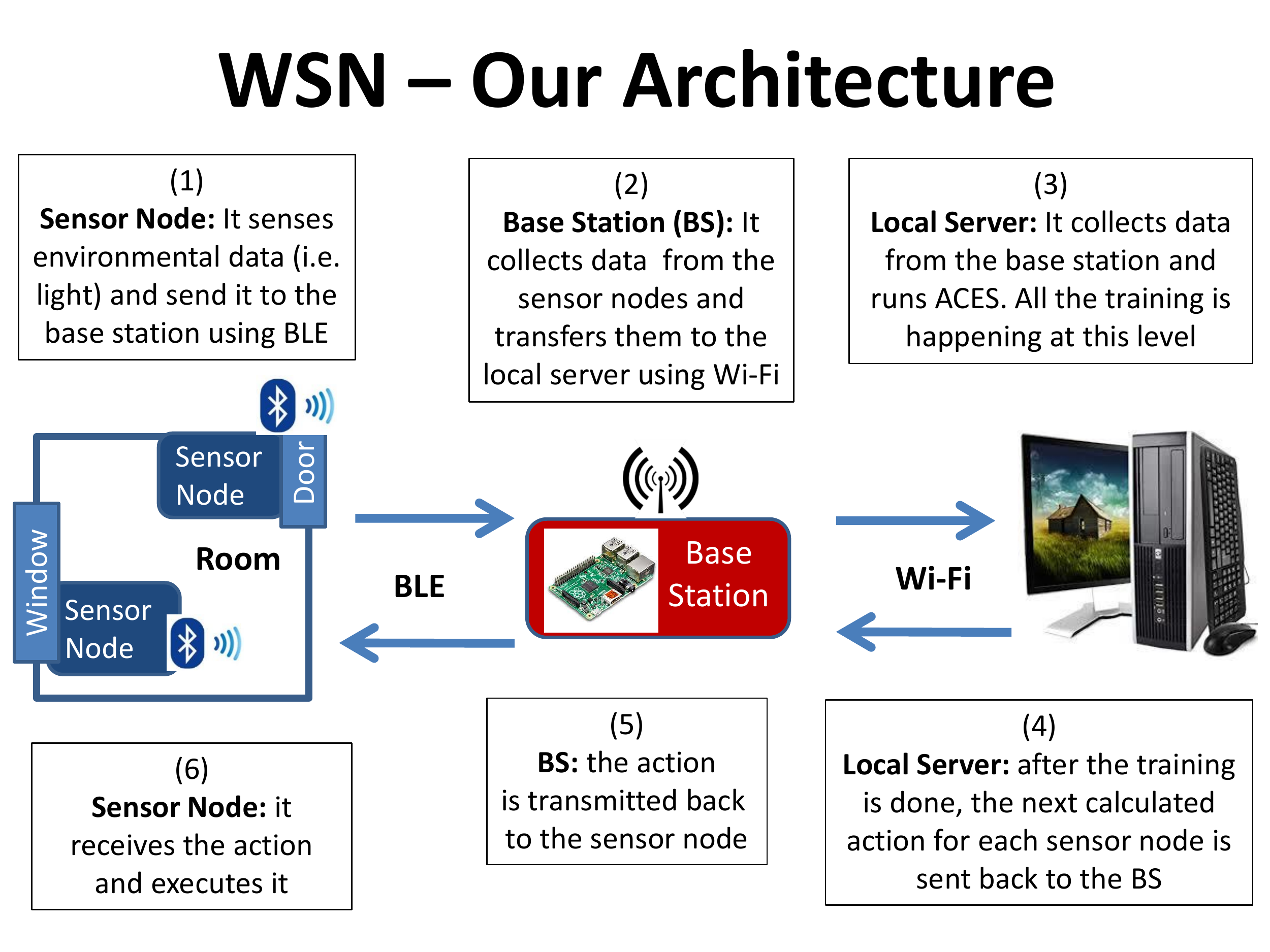}
    \caption{Our WSN architecture. The training happens at the local server and the base station is used to transfer the data from the sensor nodes to the local server and vice-versa.
    }
    \label{fig:WSN}
\end{figure}

The base station uses BLE 4.2 gattool functions to exchange data with the sensor nodes deployed around the building. As soon as a sensor-node wakes-up, it starts advertising. The base station reads the advertisement, connects to the sensor-node and exchanges data. During the connection, the base station communicates the next action to do to the sensor node while the sensor node communicates the read sensor values (i.e. light, temperature, PIR, SC Voltage) to the base station. The base station stores and sends the data to a local server for post-processing using a Wi-Fi connection. In this work, the base station is only used for passing the data from the sensor nodes to the local server. The RL training is done at the local server and the sensor nodes execute the actions decided. Figure \ref{fig:WSN} shows all the steps we just described. Due to the nature of BLE 4.2 that we use, we can only target one-hop sensor networks. \FF{Our methods can be easily extended to low bandwidth networks such as low-power wide-area networks (LPWAN). 
From \cite{mekki2019comparative} the maximum payload length for typical LPWAN protocol ranges from 8 bytes (e.g. SigFox \cite{mekki2019comparative}) to up to 1600 bytes (e.g. NB-IoT \cite{mekki2019comparative}). Therefore, our approach can be extended to LPWAN networks since our bandwidth requirements are low as transferring data of 6 bytes. We leave empirical verification for future work.}
In our system, the base station is composed of a Raspberry Pi equipped with a BLE USB dongle. In our deployment, we connected up to 15 nodes to a single base station. To facilitate large deployments, the nodes do not remain connected to the base station as in a typical Bluetooth connection but they always disconnect and reconnect using the advertisement. In this way, we are not limited by the number of simultaneous BLE connections. 


\section{Simulations}
\label{sec:eval}
We use the simulator to speed up the RL training since it can take thousands of episodes to converge. Our objective is to model the environment to sufficiently capture real-world characteristics while keeping its complexity low to allow for fast simulations. 

\subsection{Modeling the Simulation Environment}
\label{sec:env_sim}

When the agent acts on the environment, the simulation needs to respond with the next state and reward defined in Section \ref{sec:problem}. To create a simulation environment, we need to identify how much energy will be consumed and harvested under a given environment condition and sensing quality. We start with modeling the sensor platform, where we need to identify the energy consumed in different modes of operations. The energy consumption can be calculated from the datasheet of individual components used or by direct measurements in different operating modes~\cite{wang2006realistic}. We use a combination of both. For more complex platforms where individual component analysis is not feasible, we can fit a model based on power consumption in various modes of operation~\cite{rivoire2008comparison}. The energy gained is a function of both the efficiency of the harvester module and the energy available in the environment. We use raw light measurements collected by sensors in different settings to capture environment characteristics.

\subsubsection{Current Measurements:}
We use the power consumption of the solar panel and PIR from their respective data-sheets and measure the current consumption for the other components to increase the quality of our simulator.
We measure the current consumed by using the National Instrument USB-6210 with MATLAB (16-bit datum per minute). Table \ref{tab:Energy} shows the power consumption of the sensor node's main components when using a super-capacitor charged at 3V. For the sensing and transmission operations (e.g. \textit{Read Light Sensor + BLE Transmission)}, the current includes all the sensing and communications steps such as the sensor reading, the advertisement phase, the connection, the transmission of the data using BLE, and the final acknowledgment message. Therefore, we include in this value all the energy consumed by the node during its wake-up time.

\begin{table}[ht]
\centering
  \caption{Sensor-Node Current Features. For the sensing and transmission operations (e.g. \textit{Read Light Sensor + BLE Transmission}), the current includes the sensor reading, the advertisement phase, the connection and transmission of the data using BLE and the final acknowledgment message.}
  \label{tab:Energy}
  \begin{tabular}{cc}
    \toprule
    Feature & Current [\si\micro A]  \\
    \midrule
    Board Leakage + MCU in Sleep Mode & 3.5\\
    Read Light Sensor + BLE Transmission & 199 \\
    Board + PIR and MCU in Sleep Mode & 4.5\\
    PIR Detection & 102 \\
    Solar Panel at 200 lux & 31 \\
    Solar Panel at 50 lux & 7.75 \\
  \bottomrule
\end{tabular}
\end{table}

\subsubsection{Light Measurements:}
We placed a node in different types of locations and measured light intensity, supercapacitor voltage at 5-minute intervals for 15 days. The locations are: \textit{(i)} a windowless \textit{Conference Room} 
where light is On only when people occupy the room; \textit{(ii)} a \textit{Staircase} where internal lights are always On for security reasons; \textit{(iii)} the \textit{Middle of an Office} room, mainly subject to internal lights; \textit{(iv)} a node subject to natural light from a \textit{Window}; and \textit{(vi)} a node placed close to the \textit{Door} of an office room where light intensity is low.

\subsubsection{Modeling the Energy Storage Level:}
The super-capacitor accumulates energy when light is available and depletes energy with node operations.

\noindent
\underline{\textit{Energy Produced:}} 
\begin{equation}
\label{E_prod}
	E_{Prod} = Solar_{Power} * Light Intensity * Time Period
\end{equation}
 From the solar panel (i.e. AM-1454 for indoor usage) data-sheet \cite{link:sanyo}, the current generation per 200 lux of light at 2.5V is 35.2 uA. Hence, the approximate energy generated by the solar panel at 200 lux in one second is 88e-6J (35.2uA * 2.5V * 1s). As we measure light intensity only once per 5 minutes, we miss light fluctuation events. We ignore solar plan inclination, the wavelength of light and reflection of lights to keep the model simple. Hence, the energy estimated is an approximation. However, we show that the model is sufficient to learn a policy that works in the real world.  

\noindent
\underline{\textit{Energy Consumed:}} 
\begin{equation}
\label{E_cons}
	E_{Cons} = E_{Send} + E_{Sleep} 
\end{equation}
E$_{Send}$ is the energy consumed to read and transmit a sensor data packet, and E$_{Sleep}$ that is the energy consumed by the sensor node in between two transmissions. 



\subsubsection{Validation of Modeling}
\begin{figure*}[ht]
	\centering

	\includegraphics[width=0.9\linewidth]{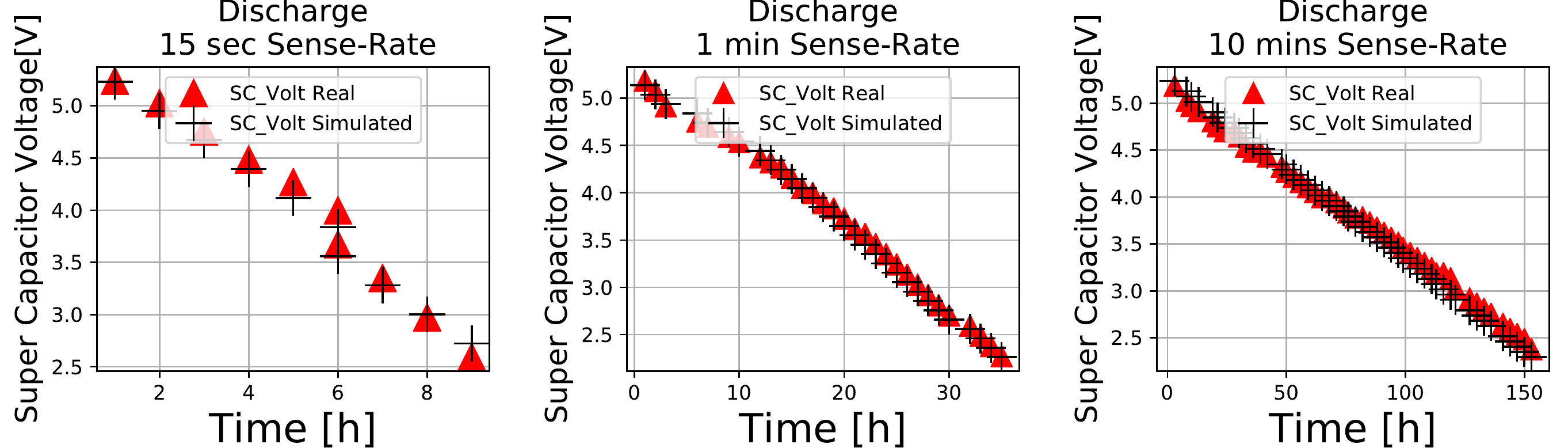}
	\caption{Super-Capacitor Discharge Comparison between our Simulator and Real-World Using Different Sensing Rates}
	\label{fig:Discharge}
\end{figure*}

We validate our discharge model when no light is available. Figure \ref{fig:Discharge} compares the lifetime of the sensor-nodes in simulations (black) and real data (red) using different sensing rates. The two trends capture the high-level characteristics well but show small differences due to real-world events that are difficult to model. 
The sensor-node lasts up to 9 hours by collecting data every 15 seconds, up to 34 hours at a 1-minute sending rate, and up to 6.25 days with a 10-minute sending-rate. 
Hence, the RL agent needs to take the right action based on the current environment state to maximize the sensing rate while avoiding energy depletion. We validate our charging model with current measurements in different light conditions as shown in Table \ref{tab:Energy}. 

\subsubsection{RL Environment Setup}
\label{sec:sim_sett}

The simulator interacts with the RL agent as explained in Algorithm \ref{algoRL}. Equations \ref{E_prod} and \ref{E_cons} keep track of the energy voltage level and the light intensity is taken from real-world measurements. We model the environment state as follows: \\
\underline{\textit{Light intensity}}: We normalize the light intensity from a range of 0 to 10, where 0 represents no light and 10 represents 2000 lux or above. We select 2000 lux as a maximum value after checking typical indoor light intensity in buildings. 

\noindent
\underline{\textit{Energy storage level}}: We scale the energy storage voltage from 0 (min voltage 2.1V) to 10 (max voltage 5.5V).

\noindent
\underline{\textit{Weekend/Weekday}}: Buildings' indoor lights patterns are strongly dependent on the presence of people \cite{paper:campbell_cinamin}. Hence, we consider a binary state to capture weekdays and weekends. 



Once the super-capacitor voltage reaches <2.1V, it terminates all its operations and energy recovery can take hours. To avoid long communication gaps between the sensor node and the base station, we penalize the RL agent with high negative reward (-300) when super-capacitor voltage is <3V.

\subsection{Simulation Results}
\label{simulation_results}
\begin{figure*}[th]
	\centering
	\includegraphics[width=\linewidth, height=7cm]{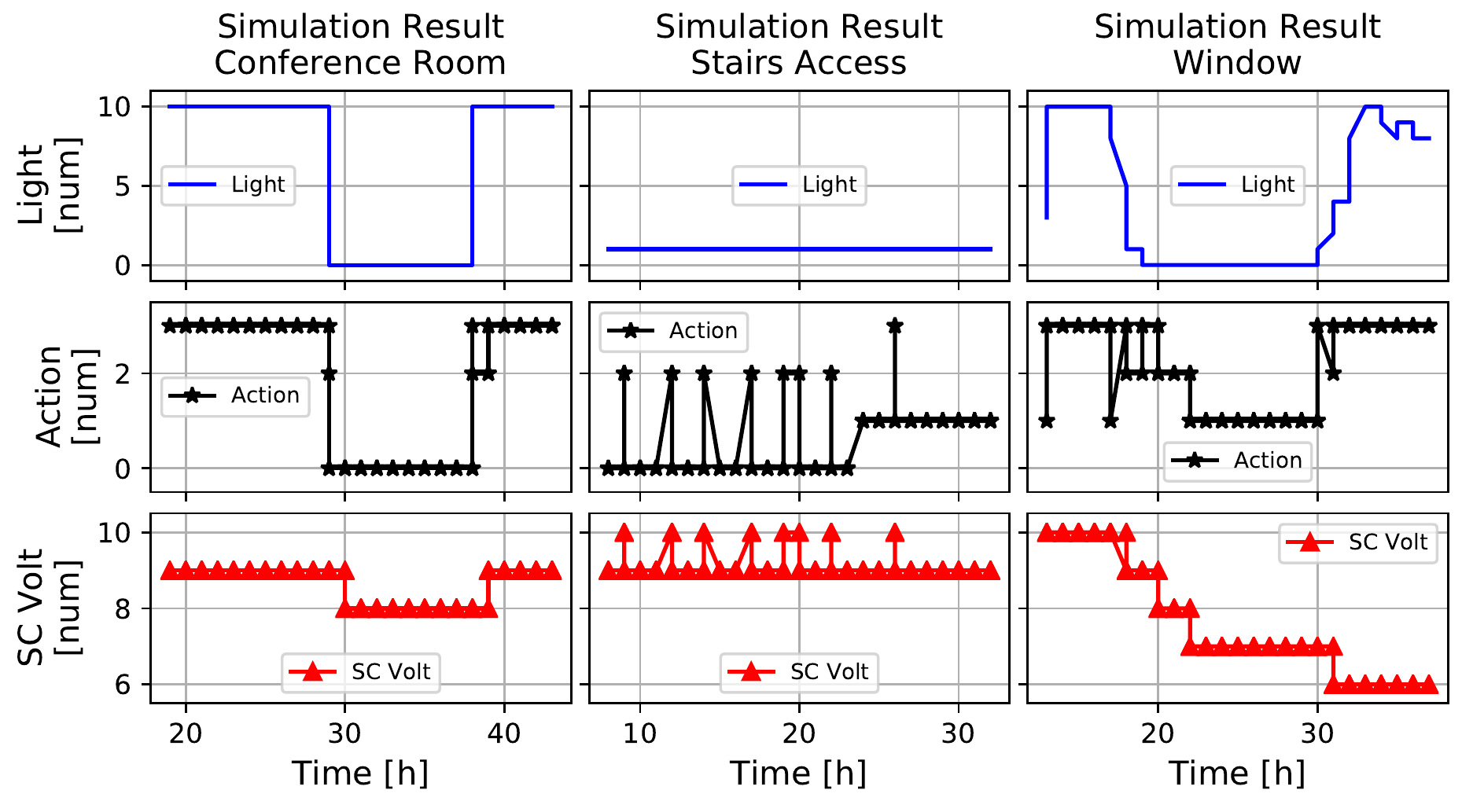}
	\caption{Simulation Result on Different Lighting Conditions. Results obtained by running ACES using 15 days of lighting data traces.
    }
	\label{fig:Simula-All}
\end{figure*}

We present the results of policies learned after deploying 5 different nodes in 5 different lighting conditions. Each node collects light measurement data every 5 minutes for 15 consecutive days. We run 5 different simulations, one per node, with the respective data collected.  Figure \ref{fig:Simula-All} shows the results of the simulation.

From Figure \ref{fig:Simula-All}-left for the Conference Room, we see that ACES uses the lowest node sleep time (i.e. action 3) when lights are on and the energy storage is almost full (SC voltage level is 9), but as soon as lights turn off, it increases the node sleep time (i.e. action 0) to save energy. The conference room has no windows, has long periods of time with lights off and hence, light patterns are sporadic. The system learns that to avoid energy storage depletion, it is better to save energy as soon as lights turn off. When the lights turn on again and the energy levels are not full, ACES 
switches between action 2 and 3 to allow the energy-storage to recover to full charge. 

From Figure \ref{fig:Simula-All}-center for the Stairs Access, we notice that light is always on at level 1 (200 lux) due to security reasons, but the intensity is not enough to keep the super-capacitor charged. Hence, ACES uses  a higher node sleep time that allows slower charging of the super-capacitor over time. When the voltage level reaches the maximum (level 10), ACES uses a lower node sleep time (action 2 or 3). That drops the super-capacitor voltage and forces ACES to use lower actions again.

From Figure \ref{fig:Simula-All}-right for the Window, we notice that ACES uses the lowest node sleep time (action 3) when the light is on and energy-storage is full, but it starts using lower actions as soon as lights go off and the super-capacitor reduces its voltage. But compared to the conference room case, the node sleep time increase is gradual as ACES learns light patterns from sunset to sunrise. It switches to a lower node sleep time even when there are no lights, forecasting that the light will become available in the next few time-steps. 

\subsubsection{Convergence Time}
\label{Convergence-time}
\begin{figure}[th]
	\centering
	\includegraphics[width=0.45\linewidth]{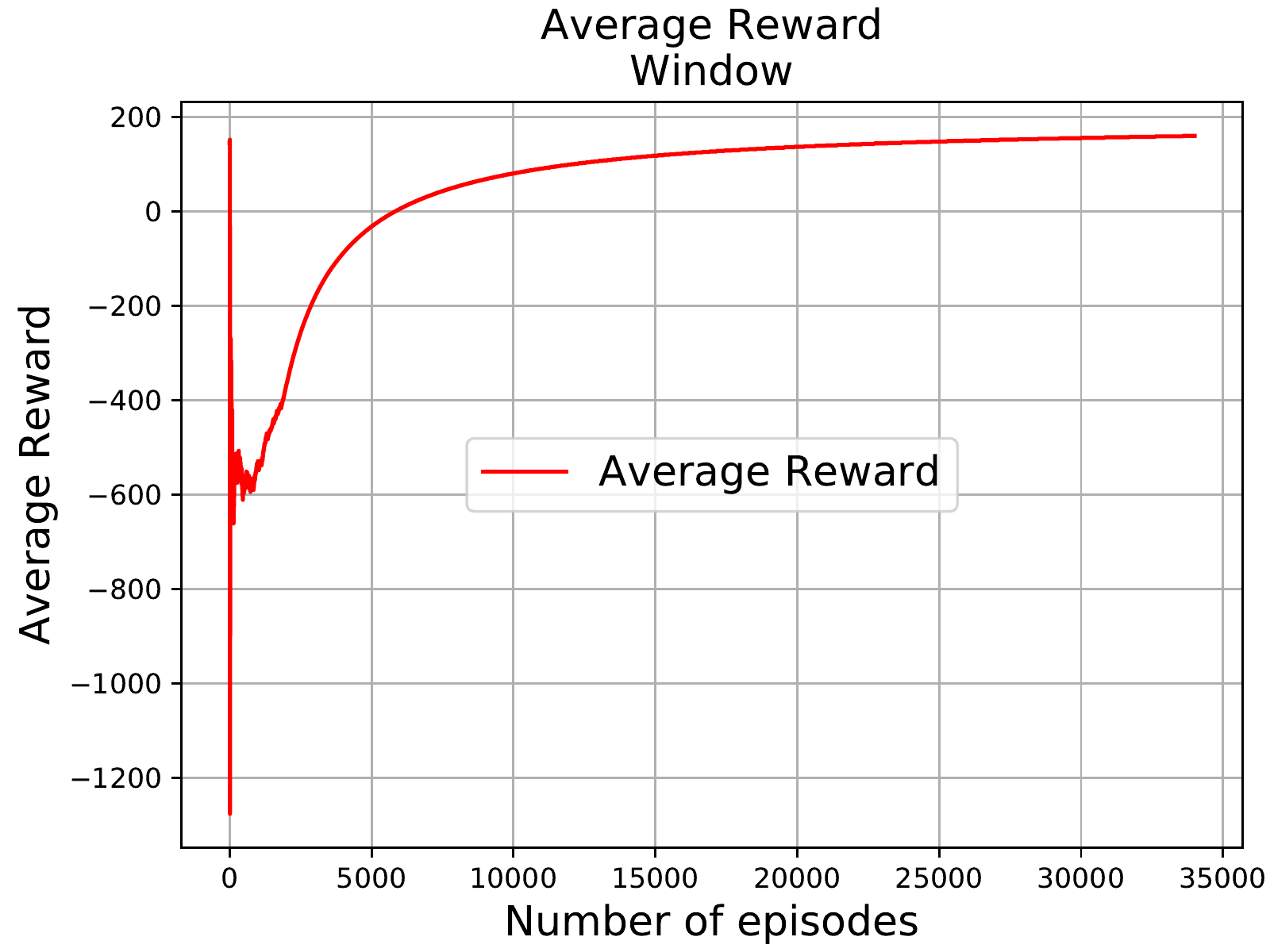}
	\caption{Average Reward Simulation Results for Window using $\alpha$ equal to 0.1. It takes less than 30 minutes for our algorithm to converge on a typical Intel Core-i7 CPU.
    }
    \label{fig:Avg_Reward-2154}
\end{figure}

For each simulation, we collect the total reward obtained at the end of each episode and average them to show the convergence of the algorithm over time.
In Figure \ref{fig:Avg_Reward-2154}, we show an example of the average reward convergence while running a simulation using $\alpha$ equal to 0.1 in the Window location. The convergence happens around 12500 episodes. 
The entire simulation takes ~30 minutes of wall-clock time using an Intel Core-i7 CPU with our Python implementation. 

\subsubsection{Input/Action Space Analysis}
\label{sec:discrete}
The dimension of the Q-Table is the product of the input and action state. By increasing the number of states: (i) we can represent the input and output variables more accurately, (ii) we need more iterations for the Q-learning algorithm to converge to a working solution.
To better understand how this trade-off behaves in our system, Tables \ref{tab:Input-Tradeoff} and \ref{tab:Action-Tradeoff} show the rewards achieved while selecting different inputs and actions respectively.

\begin{table}[ht]
\centering
  \caption{Input Space Analysis}
  \label{tab:Input-Tradeoff}
  \begin{tabular}{cccccccc}
    \toprule
    & Door & Stair & Middle & Conf & Window & Avg Rew & Input Size\\
    & [Rew] & [Rew] & [Rew] & [Rew] & [Rew] & [Rew] & [num of states]\\
    \midrule
    
    SC & 392* & 723 & 322 & 551 & 1310 & 670 & 11\\ 
    \hline
     SC-Week  & 179* & 723 & 322 & 353 & 1510 & 617 & 22\\
    \hline
    SC-Light & 135 & 699 & 527 & 611 & 1270 & 648 & 121 \\ 
    \hline
    SC-Light &  &  & & & & & \\
    Week & 161* & 721 & 832 & 1040 & 1413 & 833 & 242 \\
    \hline
    SC-Light &  &  & & & & & \\
    Week-Time & 319 & 722 & $<$0 & $<$0 & 1304 & $<$0 & 5808 \\
    \midrule
    \multicolumn{8}{ c }{SC = Super Capacitor Voltage; Time = hours of the day; Rew = Reward}\\
  \bottomrule
\end{tabular}
\end{table}

Table \ref{tab:Input-Tradeoff} reports the reward achieved by ACES when the input uses (i) the only super-capacitor voltage (i.e. SC); (ii) The \textit{SC} and \textit{Light} measure; (iii) the \textit{SC} and week/weekend day (i.e. \textit{Week}); (iv) \textit{SC, Light and Week} as for ACES and (v) \textit{SC, Light, Week} and \textit{Time} of the day that is expressed in hours. For these simulations, we use an action space equal to 4. Simulations and rewards are calculated on the same 15 days of light data as in Section \ref{simulation_results} but only the first 7 days of data are used for training. Using the only supercapacitor as an input state gives high reward on places where light is low in intensity (Door, Stair). Introducing \textit{Light} as an input state increases rewards where light has high variability throughout the day (Conference and Window). As an average of the five lighting conditions, the use of \textit{SC, Light and Week} is the one that brings more rewards to the system. The increase in rewards comes at the cost of having as an average an input space that is 22 times bigger (i.e. 242 in the Table) compared to using only the SC (i.e. 11). 
In the same Table, we conducted another experiment to understand why the reward achieved by the \textit{SC} and \textit{SC-Week} was higher than \textit{SC-Light-Week} for the \textit{Door}
case: we run the same experiment for 2.5 months of data and noticed that \textit{SC-Light-Week} achieves a higher reward in the long term (9371) w.r.t. SC-Week (9176) and SC (8186). With \textit{SC-Light-Week}, ACES uses more information to converge to a better policy, but it requires more data as it needs to learn the policy for all variations that exist in the given observations.
The use of the \textit{SC-Light-Week-Time} as an input brings the system to achieve a negative reward in 2 places. This is due to a very large input state and hence ACES could not converge to a good solution within 24 hours for which we ran the simulator. For accommodating these larger state space, we would need to use function approximation algorithms such as Deep Q-Networks~\cite{mnih2015human}. 

Table \ref{tab:Action-Tradeoff} shows the effect of changing the action space. Since the reward collected is equal to the action selected by the system at each step, for this experiment we normalized all the rewards from 0 to 3. We use \textit{SC-Light-Week} as the input state. Increasing the number of actions increases the final reward on average. We use 4 actions in our real-world experiments to keep the Q-Table small.

\begin{table}[ht]
\centering
  \caption{Action Space Analysis}
  \label{tab:Action-Tradeoff}
  \begin{tabular}{ccccccc}
    \toprule
    Reward & Door & Stair & Middle & Conf & Window & Avg\\
    \midrule
    Action = 2  &  81 & 147 & 607 & 240 & 1359 & 486\\
    \hline
    Action = 4  & 161 & 721 & 832 & 1040 & 1413 & 833 \\
    \hline
    Action = 8  & 435 & 1097 & 649 & 1235 & 1923 & 1068 \\ 
  \bottomrule
\end{tabular}
\end{table}


\subsubsection{Event-Driven Applications}
The PIR sensor adds to the sleep current of the node by 1$\mu$A, each event consumes an additional 102 $\mu$A (Table \ref{tab:Energy}) and lasts for 2.5 seconds. ACES needs to account for this additional energy use when selecting the sensing rate. We simulate stochastic events in the environment with an average of 50 events per day during weekdays and 20 events per day during the weekends, a conservative estimate of daily events as reported by Agarwal et al.~\cite{paper:yuvraj_1}. For these simulations, we use the same 15 days of data lighting traces as for Figure \ref{fig:Simula-All}, so that a comparison with previous results is fair. Again, each RL training uses lighting data collected by one node.

\begin{table}[ht]
\centering
  \caption{ACES comparison between Periodic Sensing Applications and Event-Driven Applications. Results obtained after training and running ACES on the same 15 days of data lighting traces used for Figure \ref{fig:Simula-All}.
  }
  \label{tab:Event-vs-Period}
  \begin{tabular}{cccc}
    \toprule
    Node & Periodic-Sense & Event-Driven & Percentage \\
    Placement & [avg duty-cycle period-period in sec]& [avg duty-cycle period-period in sec]&[\%] \\
    \midrule
     Conference Room & 39  & 40 & 98 \\ \hline
     Window & 30  & 31  & 96 \\ \hline
     Middle & 135 & 140  & 97 \\ \hline
     Door & 128  & 171  & 72\\ \hline
     Stairs Access & 81  & 95  & 85 \\
  \bottomrule
\end{tabular}
\end{table}

Table \ref{tab:Event-vs-Period} compares the data-packets sent by the final policy with and without the PIR sensor. In all the scenarios, ACES successfully accounts for additional energy expenditure from event-driven sensors and continues to maintain perpetual operation. 
When abundant light is available throughout the day (i.e. windows, conference room and center of office), the number of data packets sent is similar to the baseline periodic sensing (95\% to 98\%). However, in stair access and door the light availability is low, and hence, the number of data samples drops to 85\% and 72\% respectively. 

\subsubsection{Complexity of the Proposed Methods}
In Q-learning with \FF{$\epsilon$-greedy} exploration, the convergence time of the algorithm has an upper bound of O(exp(number of states)). To keep the training time tractable, we discretize the state/action space and keep the number of states to a minimum. We limit our training time to 30 minutes with a single CPU. We can scale to more number of states using deep RL techniques such as DQN \cite{mnih2015human}, which we leave for future work.
ACES uses three variants of the Q-learning algorithm: one-time learning, day-by-day learning and transfer learning. The number of states do not change between these algorithms, hence the complexity of the algorithm remains the same as Q-learning. All three versions are trained with our simulator until Q-values stabilize or 30 minutes elapse. The primary difference between these algorithms is the amount of external lighting data made available to the simulator, which in turn affects the states that are given as input to the agent. The one-time learning policy is trained only once with available data, while the day-by-day and transfer learning algorithms are trained periodically every day. The first policy trained takes as much time to converge as one-time learning. However, subsequent training converge faster as the data collected captures all the scenarios in the environment and Q-value estimates stabilize.

\begin{figure*}[ht]
	\centering

	\includegraphics[width=0.9\linewidth, height=4.5cm]{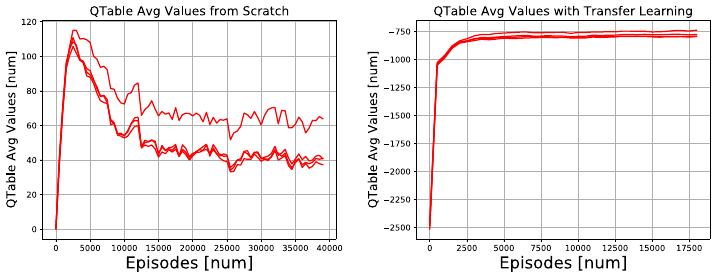}
	\caption{Convergence of the Q-Table values for one-time and transfer learning. \FF{The graphs are averaged for the actions. Therefore, we have 4 curves, one for each action. It takes 12K episodes} for Q-values to stabilize in one-time learning. The transfer learning policy takes only 3K steps to converge as it starts training from a general policy trained with data from multiple sensor nodes.}
	\label{fig:Convergence}
\end{figure*}

\FF{Figure \ref{fig:Convergence} shows the convergence} of the Q-Table values for one-time and transfer learning. \FF{The graphs are averaged for the actions. Therefore, we have 4 curves, one for each action. It takes 12K episodes for Q-values} to stabilize in one-time learning. The transfer learning policy takes only 3K episodes to converge as it starts training from a general policy trained with data from multiple sensor nodes.

\section{Real-World Experimental Results}

\subsection{One-Time Learning}
\FF{We learn policies using the simulator. Each policy trains by using 15 days of light data collected from the respective node. Then we use the resulting Q-Tables to control the real sensor nodes in the respective locations for 31 days.} We continue to update the policies based on real-world data while using $\epsilon$ = 0.9. The Q-learning updates use the parameters reported in Table \ref{tab:HyperP-OneTime} that also indicates the training-time phase and deployment-time phase.

\begin{table}[ht]
\centering
  \caption{Q-Learning Hyper-parameters for One-Time learning experiment on Real-World. The only parameter that changes compared to the simulation experiment is the $\epsilon$ that here is fixed and set to 0.9.
}
  \label{tab:HyperP-OneTime}
  \begin{tabular}{c c | c c}
    \toprule
    Hyper-Parameter & Value & Parameter & Value\\
    \midrule
    Reward-decay ($\gamma$) & 0.99 & Observations & State of Charge, \\
    & & & Light Intensity,\\ & & & Week/Weekend\\\hline
    Epsilon-fixed ($\epsilon$) & 0.9 & Actions & 15s, 60s,\\ 
    & & & 300s, 600s\\\hline
    Learning rate ($\alpha$) & 0.1 & 
    Node "Death" & 3V\\
    & & Threshold & \\\hline
    Wait time \textit{T} & 15 mins & Q-Table & $<$5\% change\\ 
    & & Convergence & in mean Q value\\\hline
    Episode Duration & 24 hours & Training-phase & 15 days\\\hline
    & & Deployment-phase & 31 days\\
    
  \bottomrule
\end{tabular}
\end{table}

Notice that the only parameter change compared to the simulation experiment is the $\epsilon$ that now is fixed and set to 0.9. Thus, the agent takes the majority of the actions based on learned Q-values from the simulator but minimally takes random actions to learn changing patterns. If the agent encounters a state not listed in the Q-table, it takes a random action.

\begin{figure*}[th]
	\centering
	\includegraphics[width=0.9\linewidth, height=6.5cm]{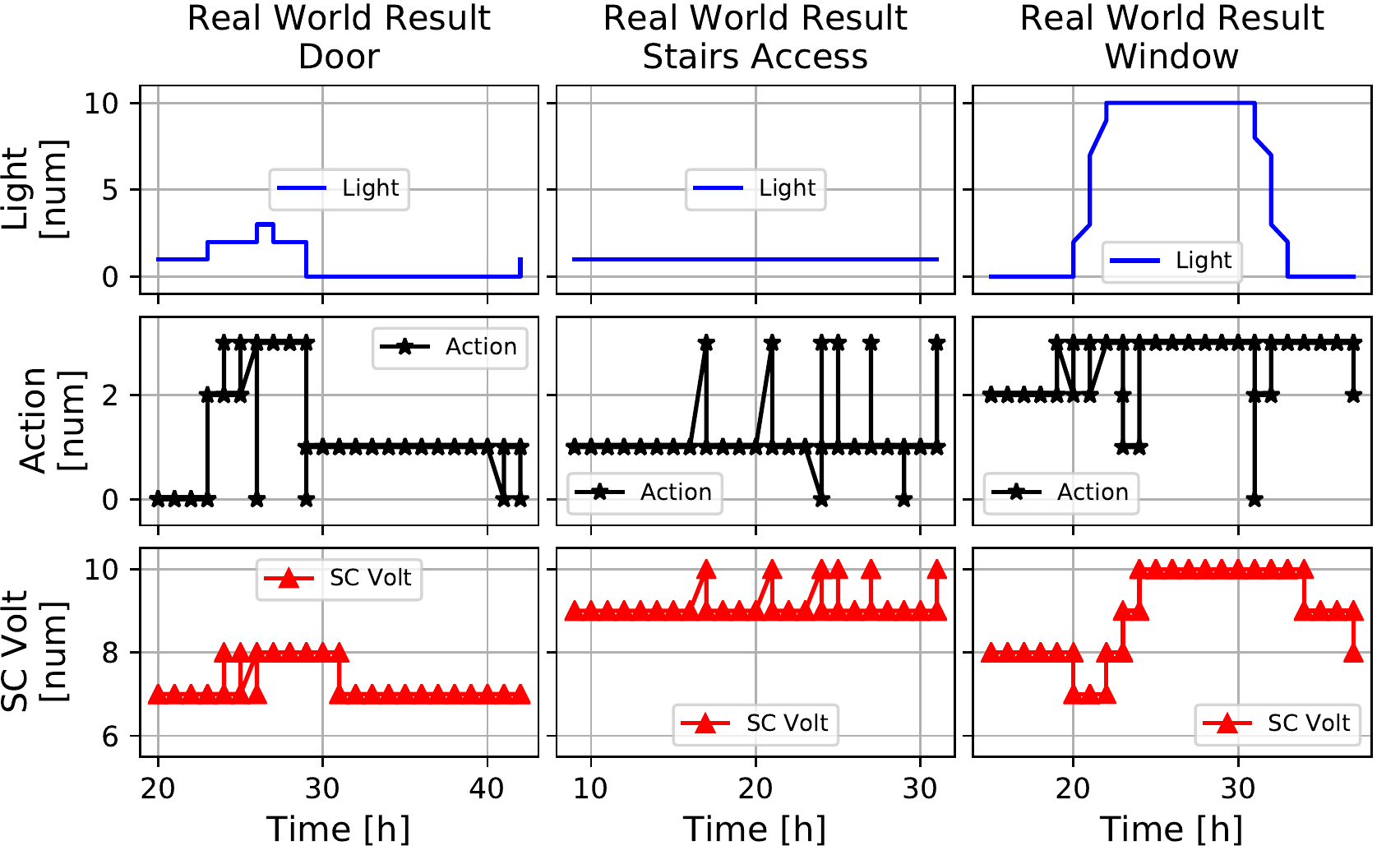}
	\caption{ACES Real-World Results in Different Lighting Conditions. As a first experiment, we use the ACES learned policies (i.e. Q-Tables) from the simulator to drive the same five nodes used to collect the 15 days lighting data traces. We leave ACES driving the nodes for 31 days. We use $\epsilon$ equal to 0.9 to leave the system exploring in the real-world. This is why we can see several outliers in the middle of a constant sequence of actions. 
    }
    \label{fig:One-Time}
\end{figure*}

Figure \ref{fig:One-Time} shows the results of three of the five nodes. The behavior in real deployment is very similar to the simulation results. There are several outliers in the sequence of actions due to $\epsilon$-greedy exploration. In the \textit{Door} location (Figure \ref{fig:One-Time}-left), the light intensity is low during the day (it reaches level 3 at most) and ACES gradually selects higher actions when enough light is available. For the Stair Access case (Figure \ref{fig:One-Time}-middle), the light availability is low, hence ACES picks lower actions (i.e. 0 or 1).

For the Window location (Figure \ref{fig:One-Time}-right), \FF{ACES uses the lowest duty-cycle period} (action 3) during the day and decreases the action to 2 when the light goes down. However, the action is never lowered to action index 1 as observed in simulations because the energy-storage level never drops to <6. Upon further digging, we found that the communication between the sensor node and the base station on average took more time than in simulation. Hence, data is exchanged less frequently (every 17 seconds instead of 15) and the power used is lower as well. The policy automatically adjusted to this difference in the environment to maximize its rewards. 

The \textit{Center of Office} results is similar to the \textit{Door} node results since it has similar light patterns. 
The \textit{Conference Room} node also performs close to the simulation results.

\subsubsection{Comparison with Fixed Periodic Sensing}

\begin{table}
\centering
  \caption{Placement-QoS comparison between a battery-powered system and ACES}
  \label{tab:Pible-vs-Batt}
  \begin{tabular}{ccccc}
    \toprule
    Node & Battery-Power & ACES & Percent & Dead \\
    Placement & [avg duty-cycle period in sec] & [avg duty-cycle period in sec] &[\%] & Time (\%)\\
    \midrule
     Conf. Room & 60 & 62 & 96 & 0\\ \hline
     Window & 60 & 35 & 170 & 0\\ \hline
     Center Office & 60 & 82 & 73 & 0\\ \hline
     Door & 60 & 136 & 44 & 0\\ \hline
     Stairs Access & 60 & 122 & 49 & 0\\
  \bottomrule
\end{tabular}
\end{table}

Table \ref{tab:Pible-vs-Batt} compares the data sent by ACES with a fixed sensing period of 1 minute used commonly in buildings. 
ACES outperforms the average duty-cycle period compared to a fixed duty-cycle period when there is a consistent amount of light as the agent learns the daylight patterns and uses a duty-cycle period of 15 seconds when the light is available (\textit{Window} case). The average duty-cycle period is similar (96\%) for the node placed in the \textit{Conference Room}. This value is closely related to the amount of light available and the presence of people in the environment. Percentages are lower for \textit{Center of Office}, \textit{Door} and \textit{Stairs Access} as the light available is lower. Even for these locations, sensor nodes send data every $\sim2$ minutes (Stairs Access: 49\%, Door: 44\%). All the 5 nodes avoided battery depletion and have 0\% dead time throughout the 31-day real-world experiment.

\subsection{Day-by-Day Learning}
\label{sec_day_by_day_results}
Instead of learning from the simulator just once with several days of data, we switch to learning a new policy from the simulator every day. Each policy is learned by using lighting data traces of the respective node. For this experiment, we deploy five nodes in five different lighting conditions for 15 days.
In our One-Time Learning deployment, we observed that by leaving the RL to explore (i.e. \FF{$\epsilon$} = 0.9) a new sequence of actions can decrease the quality of service of applications when it takes a random action once in a while. 
Hence, it is better to avoid exploration while running the sensor node in the real world. 

On the first day, we start with a fixed policy of collecting data every 15 mins. At the end of the day, the collected data is used to learn a new Q-table in the simulator by running Algorithm \ref{algoRL} with the hyper-parameters reported in Table \ref{tab:HyperP-Simu}. The learned Q-Table is used to collect data the next day with \FF{$\epsilon$} set to 1.
The hyper-parameters used to collect the data for day-by-day learning are presented in Table \ref{tab:HyperP-DbD}.

\begin{table}[ht]
\centering
 \caption{Q-Learning Hyper-parameters for Day-by-Day learning experiment on the Real-World. To avoid exploration in the real world, we set and maintain the $\epsilon$ value to 1.}

  \label{tab:HyperP-DbD}
  \begin{tabular}{c c | c c}
    \toprule
    Hyper-Parameter & Value & Parameter & Value\\
    \midrule
    Reward-decay ($\gamma$) & 0.99 & Observations & State of Charge, \\
    & & & Light Intensity,\\ & & & Week/Weekend\\\hline
    Epsilon max ($\epsilon_{max}$) & 1 & Actions & 15s,60s,\\
    \small{(For Simulation)} & & & 300s, 600s\\\hline
    Epsilon min ($\epsilon_{min}$) & 0.1 & Node "Death" & 3V\\
    \small{(For Simulation)} & & Threshold & \\\hline
    Epsilon-fixed ($\epsilon$) & 1 & Q-Table & $<$5\% change\\
    \small{(During Deployment)}& & Convergence & in mean Q value\\\hline
    Learning rate ($\alpha$) & 0.1 & Episode Duration & 24 hours \\\hline
    Epsilon increment ($\Delta$) & 0.0004  & Training-phase & repeat every 24 h \\\hline
    Wait time \textit{T} & 15 mins & Deployment-phase & 15 days\\
  \bottomrule
\end{tabular}
\end{table}

The only hyper-parameter that changes when the system takes actions in the real-world compared to the simulation experiment is the $\epsilon$ that is fixed and set to 1 during deployment.

From the second day, the training starts from the previous day's Q-table that summarizes the learning until that day. Thus, the Q-Table updates day-by-day from the first day of the deployment. For this experiment, we leave ACES driving five nodes in five different lighting conditions for 15 days.

\begin{figure}[th]
	\centering
	\includegraphics[width=0.7\linewidth]{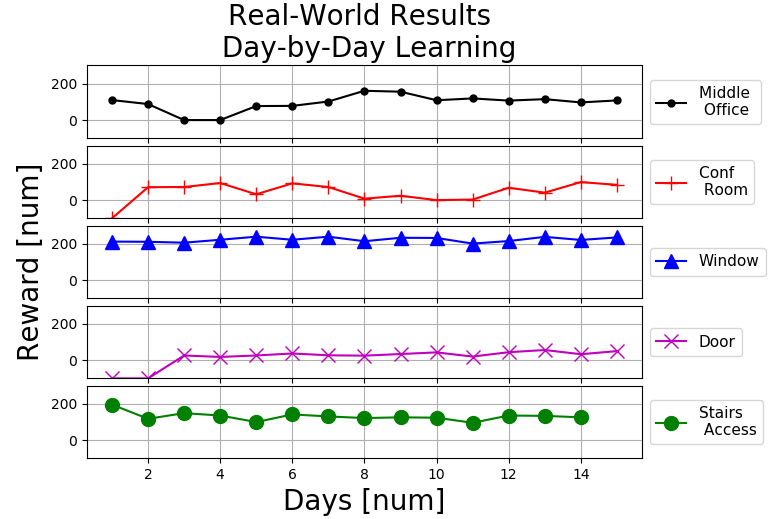}
	\caption{ACES Real-World Results in Different Lighting Conditions using Day-by-Day learning: instead of learning from the simulator just once with several days of data (i.e. 15 days in our previous experiment), we switch to learning a new policy from simulator every day. For this experiment, we leave ACES driving five nodes in five different lighting conditions for 15 days.
    }
    \label{fig:Real-Results}
\end{figure}

Figure \ref{fig:Real-Results} shows the rewards obtained by the five nodes placed on the five different lighting conditions for 15 days. The figure cuts off with a value of -10 which indicates the nodes died because of energy storage depletion. 3 out of the 5 nodes deployed (i.e. \textit{Middle Office}, \textit{Window} and \textit{Stair Access}) had 0\% dead time while two nodes (i.e. \textit{Conference Room} and\textit{ Door}) depleted their energy in the first days of the experiments. The lighting energy in the latter case is mainly subjected to human behavior (i.e. turning on lights or blinds) that can change between days. ACES needs some time to understand those patterns and adapt to them. On the other hand, the nodes that maintain continuous operations are the ones that are mainly subject to a constant light pattern - light is available from sunrise to sunset for the Window node and the light is always on for security reasons near the Stair Access node. For this experiment, we experience a dead time of 4\%.


To increase the validation of \textit{Day-by-Day} learning, we further deployed 10 more nodes for 15 days. We upgrade our boards to use a 1.5F super-capacitor to extend node lifetime in case there is no light available. The nodes achieve 0\% dead time in 15 days of deployment.





\subsection{Transfer-Learning}
We speed up the learning process for a multi-node deployment with transfer learning~\cite{taylor2009transfer}. 
\subsubsection{General Q-Table}
We use a week of lighting data collected for the day-by-day experiment (Section \ref{sec_day_by_day_results}) for all the five different lighting conditions (a total of 5 weeks of light data) to run a single ACES simulator and build a general Q-Table.
Then, we place five new sensor nodes in similar lighting conditions as the previous nodes, i.e. a node close to a window, a node in another conference room and so on. We run ACES on the new nodes while starting them with the general Q-Table instead of learning everything from scratch.

\begin{table}[ht]
\centering
  \caption{ACES results after adopting transfer learning after the first 3 days of deployment. \FF{Between parentheses}, we show the number of times the nodes died during deployment.}
  \label{tab:transferLearn}
  \begin{tabular}{cccc}
    \toprule
     Avg Duty-Cycle Period in sec & Empty & General & Percentage \\
     (Num Node Died) & Q-Table & Q-Table & [\%] \\
    \midrule
     Conference Room & 98 (1) & 64 (0) & 151 \\ 
     \textit{Avg Light [lux]} & \textit{520} & \textit{455} & \textit{89} \\ \hline
     Window & 25 (0) & 26 (0) & 98 \\ 
     \textit{Avg Light [lux]} & \textit{4518} & \textit{3888} & \textit{87} \\ \hline
     Middle Office & 101 (0) & 90 (0) & 108 \\ 
     \textit{Avg Light [lux]} & \textit{281} & \textit{340} & \textit{121} \\ \hline
     Door & 348 (2) & 301 (1) & 115\\
     \textit{Avg Light [lux]} & \textit{117} & \textit{99} & \textit{85}\\ \hline
     Stairs Access & 90 (0) & 93 (0) & 97 \\
     \textit{Avg Light [lux]} & \textit{184} & \textit{181} & \textit{98} \\
  \bottomrule
\end{tabular}
\end{table}

Table \ref{tab:transferLearn} reports the average duty-cycle period in seconds for the five different places for the first 3 days of deployment by starting ACES using a general Q-Table versus an empty one. Between parenthesis, we report the number of times the nodes died. By using a transfer learning approach, the nodes perform better as indicated by a decrease in the average duty-cycle period. The nodes subject to easy to predict patterns (\textit{Stair Access} and \textit{Window}) have similar results regardless of an empty table or a general table, ACES can find the best sequence of actions after only one day. The situation is different for the placements subject to human occupancy patterns (\textit{Conference Room}, \textit{Middle Office}, and \textit{Door}): in these cases, the general Q-Table helps speed up the learning process. The node placed in the conference room sent up to 1.5x more data.
Most importantly, the general Q-Table helps the nodes to reduce dead time: in case of the conference room, this is reduced from 1 to 0, while in the Door case this is reduced from 2 to 1.

\subsubsection{Similar Lighting Q-Table}
\label{similar_q_RL}
We also tried transfer learning by starting 5 different nodes while using a Q-Table already generated by nodes running in similar lighting conditions (e.g. \textit{Window} to \textit{Window}). All the nodes benefited from the initial policy and achieved 0\% dead time across 2 weeks of deployment. 

\subsection{Comparison with State of the Art Solutions}

We compare the average duty-cycle period for periodic sensors used by ACES using a day-by-day learning policy with \textit{(i)} an energy manager based on reinforcement learning for energy harvesting wireless sensor networks (RLMan~\cite{RLMan}), \textit{(ii)} a rule based power management algorithm that was tuned to maximize sensing-rate while avoiding energy depletion on energy-harvesting battery-less motes (i.e. Mote-Local~\cite{Pible}), \textit{(iii)} a reinforcement Q learning-based throughput on-demand provisioning dynamic power management method (RLTDPM) for sustaining perpetual operation and satisfying the ToD requirements for energy harvesting wireless sensor node \cite{paper:RL-ener-harvest-6797906}, \textit{(iv)} a dynamic sampling rate adaptation scheme based on reinforcement learning (Q-learning), able to tune sensor sampling interval on-the-fly (On-Line RL), according to environmental conditions and application requirements \cite{paper:RL-adapting} and \textit{(v)} a battery-powered system that send data every minute \cite{link:microDAQ-battery}. We consider these \FF{5 architectures} since they include current solutions (\textit{v}), literature heuristic (\textit{ii}) and RL-based methods (\textit{i, iii, iv}).

RLMan uses an actor-critic algorithm with function approximations and learns a policy based on historical data with only the energy storage level as states. We use the same state, but use the Q-learning algorithm to learn a policy. We run the algorithm until it converges in the simulator, hence the change in algorithm will not affect the final policy learned.
To simulate the Mote-Local method, we used the power management algorithm described in \cite{Pible}: the system increases the duty-cycle period if the lighting is available and the supercapacitor is increasing the voltage with time and decreases the sensing-rate when the light is off or the voltage is decreasing or maintaining the same value with time.

For this experiment, we use a week of lighting data collected for the day-by-day experiment (i.e. Section \ref{sec_day_by_day_results}).
RLTDPM is a one-time learning method. To compare it against ACES we limited the learning of the environment from the simulator to the first half of the week and let the agent take actions in the remaining half. In \cite{paper:RL-adapting} (i.e. On-Line RL), the system can only take three actions: increase, decrease or maintain the current duty-cycle period. Therefore, we modified the simulator accordingly. On-Line RL \cite{paper:RL-adapting} also uses one-time learning. Therefore, in a week-long experiment, we considered On-Line RL training using only the first half of the week data.

Table \ref{tab:Pible-vs-Batt-RLMan} shows the comparison for each lighting condition.




\begin{table}
\centering
  \caption{Average duty-cycle period (lower is better) comparison between different methods w.r.t. ACES on different indoor lighting conditions.}
  \label{tab:Pible-vs-Batt-RLMan}
  \begin{tabular}{ccccccc|c}
    \toprule
    Placement & Window & Conf. & Middle & Door & Stairs & Avg Without & ACES\\
    $[$Avg Duty-Cycle Period in sec$]$ &  & Room & Office & & Access & Door & Improvement\\
    \midrule
    Battery-Powered & 60 & 60 & 60 & 60 & 60 & 60 & + 18
    \%\\ 
    Baseline \cite{link:microDAQ-battery} & & & & & & &\\\hline\hline
    
    Mote-Local \cite{Pible} & 57 & 149 & 128 & 337 & 94 & 107 & - 34\%\\ \hline
    
     RLMan \cite{RLMan} & 30 & 133 & 150 & 411 & 136 & 112 & - 37\%\\ \hline
     
     On-Line RL \cite{paper:RL-adapting} & 69 & 165 & 103 & / & 98 & 109 & - 35\%\\ \hline
     
     RLTDPM \cite{paper:RL-ener-harvest-6797906} & 30 & 124 & 117 & / & 129 & 100 & - 29\% \\ \hline
     
     ACES & 24 & 123 & 99 & 328 & 63 & 71\\
     
  \bottomrule
  \multicolumn{8}{c}{/ = node died during experiment}\\
  \bottomrule
\end{tabular}
\end{table}

ACES outperforms the average duty-cycle period compared to the RLMan, RLTDPM, On-Line RL, and Mote-Local techniques. RL beats the Mote-Local technique as it learns the impact of each action it takes with rewards received, the formulation teaches the agent to forecast conditions based on historical data and optimize for \FF{lower duty-cycle periods}. ACES beats RLMan because it uses additional state information such as light intensity and weekday/weekend. Hence, these additional states we include in ACES makes a measurable impact on the performance of the node.

From the Stairs Access case, the duty cycle that RLTDPM can achieve is only 129 sec while ACES reaches 63 seconds. Our system obtains better rewards because of day-by-day learning. Therefore, ACES continuously learns from the environment and adapts its performance over-time even in critical low lighting conditions (e.g. Door). We put a '/' in the Door case since the node died during the simulation. On-Line RL reaches larger duty-cycle periods than ACES in every condition due to the incapacity of the system in selecting an intermediate duty cycling period when needed (i.e. Action 1 and 2). On-Line RL also depletes its energy storage in the Door case due to its inability to consider changes in the environment over-time. Our system avoids energy storage depletion thanks to day-by-day that reduces dead time to near 0.



\subsection{Real-World PIR Event + Periodical Sensing} We deployed 45 nodes with a PIR sensor to evaluate ACES in the real world for two weeks.
We place 9 nodes for each of the five lighting conditions. The nodes send both PIR events and light intensity data. 40 nodes use day-by-day learning policy and 5 nodes use transfer-learning policy. As event sensing nodes remain awake much of the time, we increased the super-capacitor size to 1.5F to account for additional energy expenditure. A node now lasts up to 9.6 hours when it sends a packet every 15 seconds with the PIR always on. It lasts up to 8 days with 10 minutes sensing period. Table \ref{tab:Event_Summary} reports a summary of the nodes deployed in the different lighting conditions, including the number of times the nodes die and the time in hours that they were off. For the \textit{Door} case, we placed two of the 9 nodes close to the ceiling near a source of light to facilitate the detection of people entering or leaving the room, and hence the average light the nodes achieves in this location is higher than other averages. 

\begin{table}[ht]
\centering
  \caption{A summary of the 45 nodes we deployed for PIR event-detection. 9 nodes deployed for each lighting condition. Numbers are average per day during a two weeks experiment.}
  \label{tab:Event_Summary}
  \begin{tabular}{ccccccc}
    \toprule
    Node & Avg & Avg & Peak & Avg & Dead & Node\\
    Placement & Light & PIR & PIR & Sensing & Time & Dead\\
     & [lux] & [event] & [event] & [sec] & [h] & [num]\\
    \midrule
     Conference & 1139 & 43 & 83 & 119 & 24 & 1\\ \hline
     Window & 4301 & 83 & 199 & 79 & 0 & 0\\ \hline
     Middle & 423 & 61 & 175 & 107 & 0 & 2$\dagger$\\ \hline
     Door & 554* & 33 & 85 & 110 & 0 & 0\\ \hline
     Stairs/ &  &  &  &  &  & \\
     Corridors & 479 & 81 & 154 & 149 & 0 & 1$\dagger$\\
     \multicolumn{7}{ c }{* 2 nodes out of 9 in the ceiling near internal lights}\\
     \multicolumn{7}{ c }{$\dagger$ 3 nodes died in the \textit{Middle} and \textit{Corridor} cases due to hardware defects}\\
  \bottomrule
\end{tabular}
\end{table}

3 nodes died in the \textit{Middle} and \textit{Corridor} cases. Upon investigation, we found that the nodes were defective and did not charge even when lights were on. For the \textit{Conference} case, most nodes performed well, but one of the nodes ran out of energy for about 24 hours. Upon checking its historical data, we found that the light in the room remained off for several days and the nodes died during the weekend. After people entered the room and turned on the light, the node resumed operations in just 15 minutes. Results are positive for the \textit{Stairs/Corridor} case where nodes detect as an average of 81 events per day and sent light data every 149 seconds. The \textit{Window} location is the most performant since sensors can send data on average every 79 seconds with 199 motion events captured.

\subsubsection{PIR Detection Accuracy}
To evaluate how many events are missed by ACES, we placed 15 battery-powered nodes as ground truth for the events detected. 
We count at most one motion event every 2 minutes for both the ACES and ground truth sensor nodes. Table \ref{tab:Event-Period-GT} compares the events detected in the different lighting conditions w.r.t to the ground truth nodes. The Table reports the average events per day.
\begin{table}[ht]
\centering
  \caption{Event detection comparison between ACES and ground-truth nodes. Events averaged per day.}
  \label{tab:Event-Period-GT}
  \begin{tabular}{cccc}
    \toprule
    Node & Ground-Truth & ACES & Percentage \\
    Placement & [events] & [events] &[\%] \\
    \midrule
     Conference Room & 52 & 48 & 91 \\ \hline
     Window & 110 & 109 & 99 \\ \hline
     Middle & 125 & 98 & 79 \\ \hline
     Door & 63 & 54 & 86\\ \hline
     Stairs/Corridors & 154 & 112 & 73 \\
  \bottomrule
\end{tabular}
\end{table}
The Table shows that for the \textit{Window} case the number of data sent from ACES w.r.t to a battery-powered node is 99\%. This is not surprising as the majority of events happen during the day. \textit{Conference} nodes also get 91 \% of the events since as soon as people enter the room, the super-capacitor gets fully charged in a few minutes. The \textit{Corridor} case is the most challenging, where there is a continuous stream of events due to people moving and ACES catches only 73\% of them (mean of 112 event per day). This can be mitigated with a different event detection strategy, e.g. reward events detected every 10 mins, for locations with large activity. 

\subsubsection{Continuous Operation and Morning First Event Detection} Barring the 3 defective sensors, 41 nodes out of the 42 deployed maintained continuous working operations. More importantly, they were operational throughout the night and detected the first PIR event in the morning when people come into their office. The first morning PIR event is important for convenience as users expect a fast response from building systems when they enter their office. ACES detects 99\% of these events and only one was delayed by 15 minutes. 

\subsubsection{Transfer Learning}
We deployed 5 of the nodes using transfer learning with a Q-Table from a node in a similar lighting condition described in Section \ref{similar_q_RL}. All the nodes benefited from the initial policy and achieved 0\% dead time.

\subsection{Energy Neutral Operations} 
It is possible that the ACES nodes perform well in our multi-week deployments, but are consuming incrementally more energy than is available and die out in a longer deployment. We evaluate the energy neutrality of ACES nodes by monitoring the super-capacitor voltage level of five randomly picked nodes in each type of location. If the super-capacitor voltage level steadily decreases over time, ACES nodes will die more frequently in longer deployments. Table \ref{tab:Energy_Neutral} shows the super-capacitor voltage percentage at midnight after two consecutive days for the five nodes. 

\begin{table}[ht]
\centering
  \caption{Energy neutral operations after consecutive days for different lighting conditions}
  \label{tab:Energy_Neutral}
  \begin{tabular}{cccc}
    \toprule
    Node & Midnight Day1 & Midnight Day2 & Difference\\
    Placement & [SC Volt in \%] & [SC Volt in \%] & [num] \\
    \midrule
     Conference &  &  &  \\ 
     Room & 93 & 100 & 7 \\ \hline
     Window & 88 & 87 & 1 \\ \hline
     Middle & 88 & 87 & 1 \\ \hline
     Door & 67 & 68 & 1 \\ \hline
     Stairs/Corridors & 91 & 91 & 0 \\
  \bottomrule
\end{tabular}
\end{table}

\begin{figure}[th]
	\centering
	\includegraphics[width=0.65\linewidth]{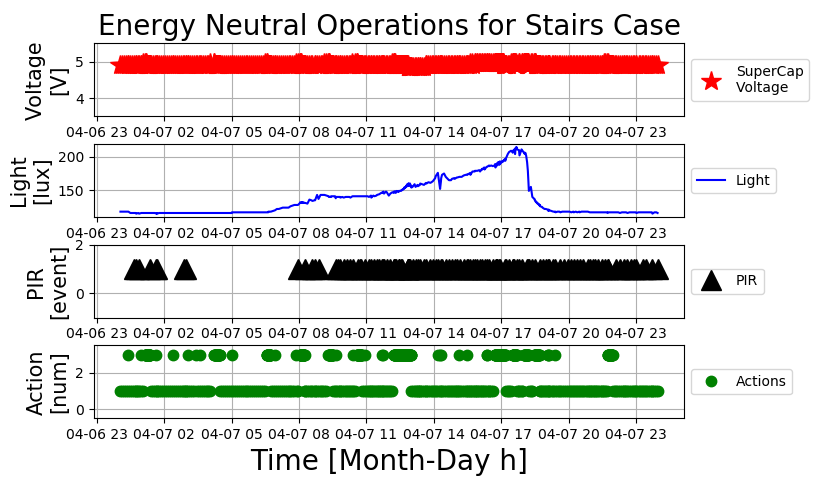}
	\caption{Energy Neutral Operations for Stairs Case. The light is almost constant throughout the day, and ACES adapts the actions to maintain a constant voltage level. At the end of two consecutive days, the supercapacitor voltage level percentage remains the same.}
    \label{fig:Energy_Neutral}
\end{figure}

For 4 out of the 5 nodes, the voltage level in the super-capacitor stays within 1\% after 24 hours. All four nodes have voltage levels at <100\%. Thus, the ACES agent learns to modulate its duty-cycle period such that it neither expends too much energy nor is it saving too much energy. Figure \ref{fig:Energy_Neutral} shows energy-neutral operations for the \textit{Stairs} Case. In this case, the light is almost constant throughout the day, and ACES adapts the action accordingly to maintain a constant voltage level. At the end of two consecutive days, the super-capacitor voltage level percentage remains the same.
 
For the Conference case, the situation is different and the supercapacitor voltage percentage reaches 100\%. In this case, the ACES agent was conservative in spending its energy and did not send as much data as it could have. The conservative strategy may work better in the Conference room because of its unpredictable changes in light conditions.

\section{Real-World Experience and Limitations}
\label{Discussion}

\begin{figure}[th]
	\centering
	\includegraphics[width=0.65\linewidth]{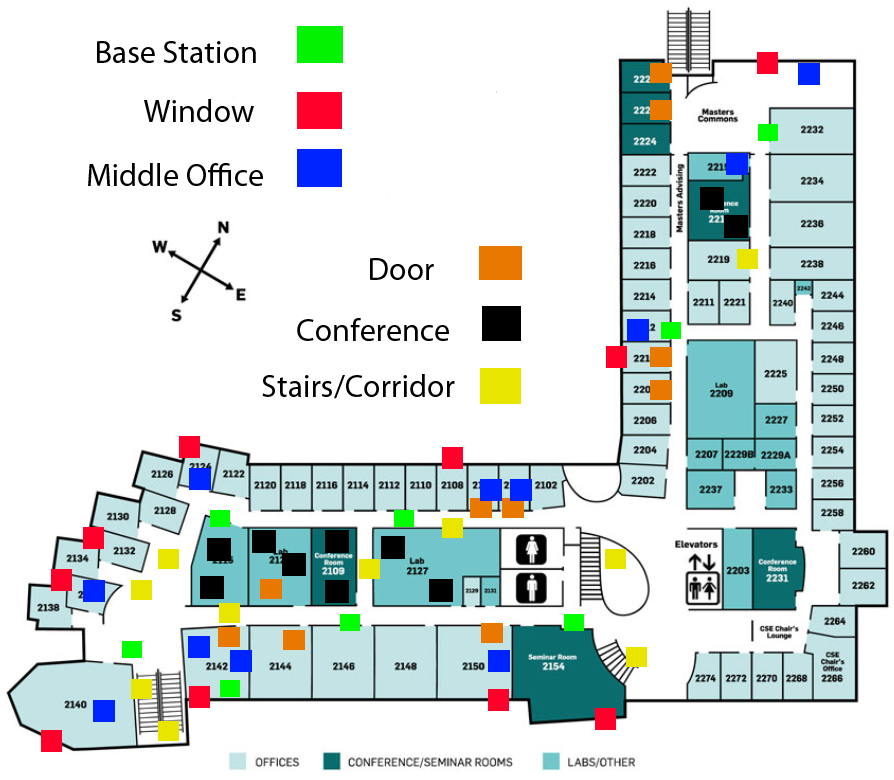}
	\caption{60 nodes deployed across our department building floor. 45 nodes send both PIR events and periodic light measurements, 15 nodes only sense light periodically. All the nodes use day-by-day learning and 10 nodes were initialized with transfer learning policy.}
    \label{fig:deployment}
\end{figure}

We have demonstrated that ACES can successfully set the sensing rate of energy harvesting devices according to light availability. Our simulation and real deployment results are promising. 
Our current deployment consists of 60 nodes across our department floor building with 15 nodes for periodical light sensing and 45 for PIR event-detection. We use 15 additional battery-powered nodes as a ground truth reference for the PIR nodes. Figure \ref{fig:deployment} shows the floor map and the position of the nodes. Based on our experience, we highlight the pertinent future research directions for using reinforcement learning at the edge. 

\underline{\textit{RL Problem Formulation:}} The key to the success of reinforcement learning is to formulate the states, actions, rewards and state transitions to capture the essence of the problem. We empirically tried several formulations before finalizing the current design. This is a one-time effort that can be applied to the deployment of numerous sensors and is much better than manually configuring each sensor. However, tools to assist domain experts in the problem specification will be immensely useful to adopt ACES like solutions.

\underline{\textit{Safe Exploration in Real-World Deployments:}} In our one-time learning policy, we kept $\epsilon=0.9$ to let the system explore during the day. Sometimes the randomly chosen action can drastically reduce the QoS. We avoided exploration by using the day-by-day learning strategy, where we learn the policy in the simulator. We can do this because of the effect of our actions on the environment were relatively easy to simulate. If that is not the case, we need a safe way to explore the real world deployment~\cite{berkenkamp2017safe}. 

\underline{\textit{Accommodate Larger State Space:}} We carefully designed our states and actions to limit the search space of the Q-learning algorithm and make the problem tractable. However, in realistic IoT deployments where nodes can have tens of parameters, we may need to accommodate large state spaces and action choices. Recently proposed deep reinforcement learning algorithms such as DQN~\cite{mnih2015human} and TRPO~\cite{schulman2015trust} can accommodate such large state spaces. As our agent resides in the power socket connected base station, it has enough computation power to deploy these algorithms. In future work, we plan to explore the computation versus performance tradeoffs among these algorithms.

\underline{\textit{Speeding-up Learning:}} In our current setup, we initially learned a new policy for each sensor node from scratch. To scale to thousands of sensors, we can learn a single policy that generalizes to many different contexts. We showed how transfer-learning can be used to speed up the learning process. However, we can learn a single policy by expanding the state space to include contextual features so that the RL agent learns the actions that maximize the long-term rewards in a context-specific manner. \FF{For our use case, we could add different states to include different information such as the type of room (conference room vs lobby), if the room has a window, etc.} A single policy can learn from data collected by all the sensor nodes and hence will also speed up learning significantly. 

\underline{\textit{Managing Network Failures:}} 
\label{sec:failures}
Communication between the nodes, the base station, and the computational unit can fail in the real-world environment due to unexpected events. During our deployment, our department network was subject to IP address reconfiguration of IP addresses by the network building manager and the local server running ACES failed to communicate the actions to the base stations and hence to the nodes. 3 nodes died during a 4-day disconnection. To limit network failures consequences, the Q-learning policy can be executed inside the node. The memory and compute requirements are low. The CC2650 MODA that is included in the board has 128Kb of programmable flash and the Q-Tables we learned are at most 25kb. As a proof of concept, we implemented a full Q-table inside one node and the node could  modulate its sensing period without the need of an external computational unit. Note that the node's computational capability is not enough for our RL training with the simulator. However, we can recover from network failures by running the Q-Table from the sensor node itself as it only takes up to 25 Kbytes of memory. 


\underline{\textit{Simulation and Real World Differences:}} While in the simulations the best sequence of actions is learned by considering all the nodes behaving identically, during our real-world deployment we discovered several factors that can differ between the nodes while severely impact the node behavior. We report them here as follows: (i) current leakage variability (ii) environmental interference, distance from the base station causing disconnections and more power needed for the node to reconnect. Taking into account those differences can increase nodes' performance. Furthermore, those differences become non-negligible in a large scale deployment. When we increased our deployment from 5 to 60 nodes, the disconnections in the network increased at a rapid rate. On average, 1 out of 20 packets is affected by disconnections and the node needs to reconnect with the base station causing increased energy consumption.

\underline{\textit{Placement of Nodes:}} One node never reached normal working functionalities because the light available was never enough for the node to charge. The node was placed in a corridor and only 60 lux was available during the day. After moving it closer to the source of light (110 lux on average), the node started working as expected. Hence, the placement of nodes should take into consideration the minimal energy required to operate the node.

\section{Conclusion}
\label{Conclusion}
Battery replacement of wireless sensors is a major bottleneck towards the adoption of large-scale deployments in buildings. We show that energy harvesting sensors provide a credible alternative when we optimize their operations using reinforcement learning (RL). Our system ACES uses RL to configure the sampling rate of solar panel based energy harvesting sensor nodes. Our sensor node uses Bluetooth Low Energy for communicating data to a nearby base station and senses both periodic measurements such as temperature and event-driven ones such as motion. Using both simulations and real-world deployments, we show that our Q-learning algorithm based RL policy learns to adapt to varying lighting conditions and sends sensor data across nights and weekends without depleting available energy. We explored deploying a one-time policy learned from historical data as well as learning a new policy based on data collected each day. Both of these strategies achieved perpetual operations in real-world deployments. We also show that transfer learning can be used to effectively \FF{reduce the initial percentage of node dead time} in real deployments.

\begin{acks}
The research reported in this paper was sponsored by the CONIX Research Center, one of six centers
in JUMP, a Semiconductor Research Corporation (SRC) program sponsored by DARPA.
\end{acks}

\bibliographystyle{ACM-Reference-Format}
\bibliography{TOSN_Main}

\end{document}